\documentstyle[12pt]{article}

\newlength{\dinwidth}
\newlength{\dinmargin}
\def\lapproxeq{\lower .7ex\hbox{$\;\stackrel{\textstyle
<}{\sim}\;$}}
\def\gapproxeq{\lower .7ex\hbox{$\;\stackrel{\textstyle
>}{\sim}\;$}}
\def\be{\begin{equation}}
\def\ee{\end{equation}}
\def\bea{\begin{eqnarray}}
\def\eea{\end{eqnarray}}

\makeatletter
\def\fmslash{\@ifnextchar[{\fmsl@sh}{\fmsl@sh[0mu]}}
\def\fmsl@sh[#1]#2{%
\mathchoice
{\@fmsl@sh\displaystyle{#1}{#2}}%
{\@fmsl@sh\textstyle{#1}{#2}}%
{\@fmsl@sh\scriptstyle{#1}{#2}}%
{\@fmsl@sh\scriptscriptstyle{#1}{#2}}}
\def\@fmsl@sh#1#2#3{\m@th\ooalign{$\hfil#1\mkern#2/\hfil$\crcr$#1
#3$}}
\makeatother

\begin{document}
\titlepage
\begin{flushright}
DTP/97/14  \\
March 1997 \\
\end{flushright}

\begin{center}
\vspace*{2cm}
{\Large \bf Higgs studies in polarized $\gamma\gamma$ collisions}
\\
\vspace*{1cm}
V.S.\ Fadin$^{1,2}$, V.A.\ Khoze$^{1,3}$ and A.D.\ Martin$^1$
\end{center}

\vspace*{0.5cm}
\begin{tabbing}
$^1$xxxx \= \kill
\indent $^1$ \> Department of Physics, University of Durham,
Durham, 
DH1 3LE, UK. \\

\indent $^2$ \> Budker Institute for Nuclear Physics and
Novosibirsk State 
University, \\
\> 630090 Novosibirsk, Russia. \\

\indent $^3$ \> INFN - Laboratori Nazionali di Frascati,
PO Box 13, 
00044, Frascati, Italy.
\end{tabbing}

\vspace*{2cm}

\begin{abstract}
The study of an intermediate mass Higgs boson, via the process
$\gamma\gamma 
\rightarrow H \rightarrow b\overline{b}$ from an initially
polarized $J_z = 0$ 
state, has been advocated as an important feasible goal of a
future photon 
linear collider.  The crucial argument was the $m_b^2/s$
suppression of the 
background process $\gamma\gamma (J_z = 0) \rightarrow
b\overline{b}$. We critically review the contribution of the
radiative background processes (in which the $m_b^2 / s$ suppression
is absent) to the quasi-two-jet-like events with at least one, but preferably 
two, tagged energetic $b$ jets. Within a complete study of the radiative
processes, we find that a
sizeable background contribution can come from the helicity-violating 
$\gamma \gamma (J_z = 0) \rightarrow b \overline{b}$ 
process accompanied by soft gluon emission.  These
latter radiative corrections contain a new type of double logarithmic (DL) terms.
We clarify the physical nature of these novel DL corrections.  Despite the fact
that the one-loop DL terms are comparable or even larger than the Born term, 
fortunately we find that the calculation of the cross section in the two-loop approximation is
sufficient for a reliable evaluation of the background to the 
Higgs signal. 
\end{abstract}

\newpage

\noindent {\large \bf 1.~~Introduction}

Now that the top quark has been discovered, the Higgs particle
$H$ is the only 
fundamental object of the Standard Model which has not been found
experimentally.  
Many theoretical studies have been performed (see, for example,
reviews 
\cite{GHKD}-\cite{CZA}) in order to examine various aspects of
Higgs hunting.  
Searches in the near future are concentrating on the possibility
of finding a 
Higgs boson in the so-called intermediate mass region,
\be
65 \; \lapproxeq \; M_H \; \lapproxeq \; 140 \: \hbox{GeV}.
\label{eq:a1}
\ee
Within the Standard Model such a Higgs particle decays dominantly
into a 
$b\overline{b}$ pair with the decay coupling being proportional
to the $b$-quark 
mass $m_b$.

In this connection it is relevant to note that various
fundamental physics 
issues could be examined in the collisions of high-brightness,
high-energy 
photon beams at future linear colliders (see e.g.\
\cite{GB}-\cite{VAK1}).  
In fact the rapid advances of laser technology make possible just
such a new 
type of experimental facility known as a Photon Linear Collider
or PLC 
\cite{IFG2}-\cite{VIT} in which high energy photon beams are
produced by the 
Compton back-scattering of laser photons off linac electrons. 

One particularly 
interesting use of the PLC would be to measure the two-photon
decay 
width of a Higgs boson once it is discovered \cite{GH,BBC}.  The
$\gamma \gamma$ 
width of $H$ is one of its most important properties.  The coupling 
of the Higgs to two photons proceeds through a sum of loop
diagrams for all 
charged particles which couple to the Higgs.  For example, the
decay width 
$\Gamma (H \rightarrow \gamma\gamma)$ can explore the possible
existence of quarks heavier than the top since they contribute
without
being suppressed 
by their large mass.  Therefore this channel may provide a way to
count 
the number of such heavy quarks.

In a PLC, the partial width $\Gamma (H \rightarrow \gamma\gamma)$
is deduced by 
measuring the Higgs production cross section in the reaction
\be
\gamma \gamma \; \rightarrow \; H \; \rightarrow \;
b\overline{b}.
\label{eq:a2}
\ee
The number of detected events is proportional to the product
$\Gamma (H 
\rightarrow \gamma\gamma) \: B (H \rightarrow b\overline{b})$. 
Thus, a 
measurement of the 
$b\overline{b}$ production cross section can, in principle,
determine this 
product.  An independent measurement of the branching ratio $B (H
\rightarrow 
b\overline{b})$, say at an $e^+ e^-$ 
collider in the process $e^+ e^- \rightarrow ZH \rightarrow ZX$
\cite{EGN,IK}, 
then allows a determination of the $\gamma\gamma$ partial width. 
However, to 
isolate $b\overline{b}$ production induced by an intermediate
mass Higgs boson 
we must first suppress the continuum 
\be
\gamma \gamma \; \rightarrow \; q\overline{q} \quad (\hbox{with}
\; q \: = 
\: b,c)
\label{eq:a3}
\ee
background events which lie beneath the resonant signal
$(\gamma\gamma 
\rightarrow H \rightarrow b\overline{b})$, assuming that the $b$
and $c$ quarks 
can be distinguished from light quarks by tagging of at least one
heavy quark 
jet.

In order to suppress the continuum background it has been
proposed \cite{GH,BBC} 
that we exploit\footnote{We assume that it is experimentally
possible to separate $J_z = 0$ and $|J_z| = 2$ $\gamma\gamma$ 
beams.  The $z$ axis is taken along one of the incoming photon
beam 
directions.  According to the present 
understanding it appears feasible to achieve a polarisation ratio
$P = (J_z = 0)/(|J_z| = 2)$ of 20--50 at a PLC \cite{VIT,BBC}.}
the
polarisation dependence of the $\gamma\gamma \rightarrow 
q\overline{q}$ cross sections (e.g.\ \cite{INOK,TB}).  Recall
that the Higgs 
signal is produced from a $\gamma\gamma$ initial state with $J_z
= 0$.  The idea 
is that the background is dominantly produced from a $J_z = \pm
2$ initial 
state, whereas the $J_z = 0$ (Born) cross section is suppressed
for large angle $q$ and $\overline{q}$ production by a factor of
$m_q^2/s$
\cite{GH,INOK,BKSO}.
  
The physical origin of this suppression \cite{VAK1,BKSO} is
related to the symmetry properties of the helicity amplitudes
$M_{\lambda_1, 
\lambda_2}^{h,\overline{h}}$ describing the background process
\be
\gamma (\lambda_1, k_1) \: + \: \gamma (\lambda_2, k_2) \;
\rightarrow \; q 
(h, p) \: + \: \overline{q} (\overline{h}, \overline{p}).
\label{eq:a4}
\ee
Here $\lambda_i$ are the helicities of the incoming photons, and
$h$ and $\overline{h}$ are the (doubled) helicities of the
produced quark
and antiquark.  The $k$'s and $p$'s denote the particle
four-momenta.  It 
can be shown, using an analogous argument to that in
\cite{BKSO}, that the 
real part of the amplitude for a $J_z = 0$ initial state
$(\lambda_1 = \lambda_2)$ 
and the opposite helicities of the quark and antiquark
$(\overline{h} = -h)$ 
vanishes in all orders in perturbative theory, that is
\be
{\rm Re} \: M_{\lambda, \lambda}^{h, -h} \; = \; 0.
\label{eq:a5}
\ee
If we take into account quark helicity conservation, then for
large angle 
production we also have
\be
M_{\lambda, \lambda}^{h,h} \; \sim \; {\cal O} \left (
\frac{m_q}{\sqrt{s}} 
\right ) \: M_{\lambda, - \lambda}^{h, - h},
\label{eq:a6}
\ee
where the amplitude on the right-hand-side displays the dominant
helicity configuration of the background process at large 
angles.  The above-mentioned $m_q^2$ suppression of the $J_z = 0$
Born cross section is a consequence of Eqs.\ (\ref{eq:a5}) and
(\ref{eq:a6}) 
and the fact that the Born amplitudes are real.  Note that we are
only concerned with
background 
$\gamma\gamma \rightarrow q\overline{q}$ production at large
angles since this is the event topology of the Higgs
signal\footnote{Here 
we require that most of the $\gamma\gamma$ collision energy is
deposited in the central detector.  This provides a very strong 
suppression of the resolved 
photon contributions, such as $\gamma \rightarrow gX$, followed
by $g\gamma \rightarrow q\overline{q}$ \cite{BKSO,BBB}.  Such
processes will therefore not be considered here.}. 
Here and in 
what follows we will take the $\gamma\gamma$
centre-of-mass collision
energy $\sqrt{s} = M_H$.

In the Born approximation the straightforward calculation of the
above $J_z = 0$ 
amplitudes gives
\be
\left ( M_{\lambda, \lambda}^{h, \overline{h}} \right )_{\rm
Born} \; = \; 
\frac{8 \pi \: \alpha \: Q_q^2}{(1 - \beta^2 \: \cos^2 \:
\theta)} \:  
\frac{2 m_q}{\sqrt{s}} \; (\lambda \: + \: \beta h) \: \delta_{h,
\overline{h}} ,
\label{eq:a7}
\ee
where $\beta \equiv \sqrt{1 - 4 m_q^2/s}$, and $m_q$ and $Q_q$
are the mass and 
electric charge of the quark respectively.  Thus we see that in
the Born 
approximation $\gamma\gamma \rightarrow q\overline{q}$ production
in the $J_z = 
0$ channel is suppressed\footnote{We see that in the high energy 
limit the $J_z = 0$ amplitude is additionally $m_q$-suppressed
when $\lambda = 
-h$.  This suppression is readily seen in the results of ref.\
\cite{INOK}.} by 
a factor $m_q^2/M_H^2$.  That is for at energies $\sqrt{s}
\approx M_H$
\be
d \sigma^{\rm Born} \: (J_z = 0) \; \sim \; \frac{m_q^2}{M_H^2}
\; d 
\sigma^{\rm Born} \: (J_z = \pm 2)
\label{eq:a8}
\ee
with the $J_z = \pm 2$ cross section having the normal behaviour
\be
d \sigma^{\rm Born} \: (J_z = \pm 2) \; \sim \;
\frac{\alpha^2}{M_H^2}.
\label{eq:a9}
\ee
So far so good --- in the $J_z = 0$ $\gamma\gamma$ channel the
$b\overline{b}$ 
$(c\overline{c})$ background process appears to be suppressed by
a large factor, 
$m_q^2/M_H^2$.  However, we must consider contributions beyond
the Born 
approximation.

First, we note that the amplitude $M_{\lambda, \lambda}^{h, - h}$
acquires an imaginary part related to discontinuities of diagrams
of the type shown in Fig.~1.  The corresponding ${\cal O}
(\alpha_S^2)$ contribution to the $\gamma\gamma \rightarrow
q\overline{q}$ cross section, $d \sigma^{\rm Im}$, is
non-zero\footnote{It vanishes in the special case of scattering
at $\theta = 90^\circ$ \cite{VAK1,BKSO}.} in the $m_q = 0$ limit.
In fact, at very high energies this contribution dominates large
angle $q\overline{q}$ production from the $J_z = 0$ initial
state.  However, an explicit calculation \cite{JT1} shows  (in
the central region, $\theta \sim 1$, and for energy $\sqrt{s} =
M_H \sim 100$ GeV) that $d \sigma^{\rm Im}$ $(J_z = 0)$ is at
least an order of magnitude smaller than the Born approximation
result. The ratio reaches its maximum at $\theta = \pi/4$ where
\be
\left ( \frac{d \sigma^{\rm Im} \: (J_z = 0)}{d \sigma^{\rm Born}
\: (J_z
= 0)} \right )_{\rm max} \; 
\simeq \; \frac{2}{9} \: \frac{\alpha_S^2 (M_H^2)}{\pi^2} \:
\frac{M_H^2}{m_b^2} \; \lapproxeq \; 0.1,
\label{eq:a10}
\ee
for $M_H \approx 100$ GeV.  We therefore neglect the $d
\sigma^{\rm 
Im}$ contribution from now on.

More seriously, the $m_q^2/M_H^2$ suppression of the $J_z = 0$
$\gamma\gamma 
\rightarrow q\overline{q}$ background is, in principle, removed
by gluon 
bremsstrahlung in the final state \cite{BKSO}\footnote{The impact
of the radiative 
background processes on the phenomenology of an intermediate mass
Higgs boson 
at PLC has also been studied in the recent papers
\cite{BBB}-\cite{KMC}.}.  In 
other words the radiative process $\gamma\gamma \rightarrow
q\overline{q}g$ 
(with $q = b,c$) can have a dramatic effect.  It can mimic the
$b\overline{b}$ 
two-jet topology of the Higgs signal in two important ways:  (i)
if partons 
are quasi-collinear, for example, a fast quark recoiling against 
a collinear quark and gluon, or (ii) if one of the partons is
either quite soft 
or is directed down the beampipe and is therefore not tagged as a
distinct jet.  
A particularly interesting example \cite{BKSO} of the latter is
when one of the 
incoming photons splits into a quark and an antiquark, one of
which carries 
most of the photon's momentum and Compton scatters off the other
photon, 
$q (\overline{q}) \gamma \rightarrow q (\overline{q}) g$ (see
Fig.\ 2).  Two 
jets are then identified in the detector, with the third jet
remaining 
undetected.

In the ideal situation in which we clearly identify two narrow
$b$ quark jets, 
the radiative background is not a problem.  However, in the
realistic experimental 
situation the isolation of the Higgs signal will be much more
problematic.  To 
reduce the (radiative) background it will be necessary to perform
a detailed 
study of the optimum jet shape cuts and to consider the
efficiency of the 
separation of $b$ jets from $c$ jets (see, for example,
\cite{BBC,BKSO,OTW}).  In 
fact distinguishing $b$ from $c$ jets will be a crucial
experimental task.  We 
note the factor of 16 amplification of $\gamma\gamma \rightarrow
c\overline{c}$ 
over $\gamma\gamma \rightarrow b\overline{b}$ due to the
different charges of 
the quarks.

Our main concern here is the calculation of the radiative
corrections to the 
background process $\gamma\gamma \rightarrow q\overline{q}$,
which is found to 
have several interesting features in its own right. We begin our study 
with a brief review of how the radiative 3 jet process, 
$\gamma\gamma \rightarrow q\overline{q}g$, can mimic the Higgs
$\gamma\gamma 
\rightarrow H \rightarrow b\overline{b}$ signal in the so-called collinear
and Compton configurations.  In Section 3 we extend the existing evaluation
of the Compton contribution to the case of polarised photons using the method
of quasi-real fermions.  We then turn from radiative to non-radiative 
corrections.  In Section 4 we clarify the physical origin of the novel
non-Sudakov double logarithmic (DL) terms which occur in helicity-violating
amplitudes.  The DL terms look particularly dangerous for $\gamma \gamma
\rightarrow q \overline{q}$ amplitudes with equal photon helicities because
of the large coefficient at the one-loop level (namely $c_1 = -8$ in
(\ref{eq:a21}) and (\ref{eq:i6})).  To investigate their total potential 
effect we therefore, in Section 5, calculate the DL contribution at the two-loop
level.  Fortunately, we find that this is sufficient to provide a reliable
evaluation of the non-radiative corrections.  We explain the physical reason 
why this is so.  In Section 6 we finish with a short discussion of the impact
of radiative corrections on the background to the Higgs signal that can be
produced in polarised photon-photon collisions. \\

\noindent {\large \bf 2.~~Overview of radiative corrections}

The non-radiative backgrounds of the Higgs signal in
$\gamma\gamma$ collisions were considered in \cite{BBC}.  With
highly 
polarized photon beams, such backgrounds were believed to be
small and 
hence thought not to hinder the study of an intermediate-mass
Higgs boson 
at a $\gamma\gamma$ collider.  For example, if the resolution for
reconstructing 
the Higgs mass $M_H$ is 10 GeV then in the Born approximation the
ratio $R$ of 
the signal to background $b\overline{b}$ events in the $J_z = 0$
channel is
\be
R \; \sim \; \left ( \frac{M_H}{60 \: {\rm GeV}} \right )^5.
\label{eq:b10}
\ee
This estimate applies if the Higgs lies in the 65--120 GeV mass
interval, see Refs.\ \cite{BBC,BKSO}.

The analysis 
of \cite{BBC} was based only on Born level calculations of
$b\overline{b}$ (and 
$c\overline{c}$) production.  However, more recently it has been
pointed out 
\cite{BKSO}--\cite{KMC} that QCD corrections are 
very important and considerably complicate the extraction of the
Higgs signal 
from the background.  The problem is evident once we note that
the cross section 
for the radiative background process $\gamma\gamma \rightarrow
q\overline{q}g$, unlike the Born signal of (\ref{eq:a8}), does
not contain the
$m_q^2/M_H^2$ suppression factor.

The aim of this paper is to systematically study the QCD
radiative corrections 
to $\gamma\gamma \rightarrow q\overline{q}$ in the $J_z = 0$
channel.  Before 
we present detailed calculations it is informative to give
order-of-magnitude estimates of the relevant processes.  We have
to consider 
corrections to the two-body $b\overline{b}$ (or $c\overline{c}$)
final state, 
as well as studying the impact of radiative $q\overline{q}g$
production.  As 
mentioned above, the latter is a particular problem for the
$\gamma\gamma 
\rightarrow H \rightarrow b\overline{b}$ jet signal in two
different kinematic 
regimes:  the collinear and Compton configurations.

By the collinear configuration we mean the production of the $q$
and 
$\overline{q}$ at large angles accompanied by gluon radiation
with limited 
maximal energy $\Delta E_g$ in the direction orthogonal to the
most energetic quark.  This applies in the region
\be
\frac{m_q}{M_H} \; < \; \varepsilon_g \; \ll \; 1,
\label{eq:a11}
\ee
where $\varepsilon_g = \Delta E_g/M_H$.  Then we have
\be
\sigma \left ( \gamma\gamma|_{J_z = 0}\rightarrow q\overline{q}g
\right ) \; 
\sim \; \frac{\alpha^2 Q_q^4}{M_H^2} \; \frac{\alpha_S
(\varepsilon_g
M_H)}{\pi} \; 
\varepsilon_g \: \ln \frac{1}{\varepsilon_g},
\label{eq:a12}
\ee
which manifestly does not contain the factor $m_q^2/M_H^2$, but
which can be 
suppressed by taking $\varepsilon_g$ small.  If the acollinearity
of the $q$ 
and $\overline{q}$ jets were allowed to be large 
such that $\varepsilon_g \sim {\cal O} (1)$ then the radiative
background 
becomes
\be
\sigma \left (\gamma\gamma |_{J_z = 0} \rightarrow q\overline{q}g
\rightarrow 3 \: 
{\rm jets} \right ) \; \sim \; \sigma \left (\gamma\gamma |_{J_z
= \pm 2} 
\rightarrow q\overline{q}g \rightarrow 3 \: {\rm jets} \right )
\; \sim \; 
\frac{\alpha^2 Q_q^4}{M_H^2} \; \frac{\alpha_S (M_H)}{\pi}
\label{eq:a16}
\ee
which greatly exceeds the Higgs signal --- so clearly this is 
not a regime in which to search for the Higgs boson.

The Compton configuration of the $q\overline{q}g$ background is
shown in Fig.~2.  
In this case it is the outgoing quark (or antiquark) along the
photon beam 
direction which is comparatively soft, that is
\be
\frac{m_q}{M_H} \; < \; \varepsilon_q \; \ll \; 1,
\label{eq:a14}
\ee
where $\Delta E_q = \varepsilon_q M_H$ is its maximal allowed
quark energy.  The 
cross section is of the form
\be
\sigma \left ( \gamma\gamma |_{J_z = 0, \pm 2} \rightarrow 
q\overline{q}g \right )_{\rm Compton} \; \sim \;
\frac{\alpha^2 Q_q^4}{M_H^2} \; 
\frac{\alpha_S (M_H)}{\pi} \; \varepsilon_q \: \ln \: 
\frac{\varepsilon_q M_H}{m_q},
\label{eq:a13}
\ee
where the logarithm arises from the integration over the
transverse 
momentum of the spectator quark.  Incidentally, to separate the
dominant 
contribution (\ref{eq:a13}) arising from 
one energetic and one comparatively soft $b$ quark from the Higgs
topology with two fast $b$ quarks may pose an experimental
challenge, see
Ref.\ \cite{BKSO} for details.

In summary, to have a chance to isolate the Higgs contribution we
require the 
observation of two energetic $(b,\overline{b})$ jets with at
least one, and 
preferably two, tagged $b$ quarks.  To reduce the radiative
$q\overline{q}g$ 
background we must impose kinematic cuts, such as $\varepsilon_g,
\varepsilon_q \ll 1$.  
However, there is a price to pay, since the signal is also
depleted.  Indeed 
\be
\sigma \left (\gamma\gamma \rightarrow H \rightarrow
b\overline{b} \; {\rm jets} 
\right ) \; = \; \tilde{\sigma} \left ( \gamma\gamma \rightarrow
H \rightarrow 
b\overline{b} \right ) \: F_g.
\label{eq:a17}
\ee
where $F_g$ is a Sudakov form factor \cite{VVS} which occurs
because we need to 
impose a cut (say $\varepsilon_g \ll 1$) in order to prohibit
energetic gluon 
emissions.  $F_g$ involves a resummation of double logarithmic 
terms.  Its explicit form depends on the cut-off conditions on
the 
accompanying gluon radiation, see for details the end of section
4.  
Thus when imposing the restriction (\ref{eq:a11}) we need to
resum 
the $(\alpha_S L_g^2)^n$ terms where
\be
L_g \; \equiv \; \ln \left ( \frac{1}{\varepsilon_g} \right ).
\label{eq:b17}
\ee
If, on the other hand, the gluon energy is restricted, $k_{\rm
max} = 
k_0$, then a resummation of DL terms of the form $(\alpha_S L_m
L_0)^2$ 
is required.  Here 
\be
L_m \; \equiv \; \ln \left ( \frac{M_H}{m_q} \right ), \quad\quad
L_0 \equiv \ln \frac{M_H}{k_0}.
\label{eq:a18}
\ee
The form factor decreases rapidly if $\varepsilon_g$ (or $k_0$)
is taken to
be smaller.  
That is the more we require the $b$ and $\overline{b}$ jets to be
collinear 
then the more we reduce the signal.  The second effect, which
introduces 
$\tilde{\sigma}$ in (\ref{eq:a17}), is due to higher order
(single 
logarithmic, $ \alpha_S L_m$) QCD effects which are known
\cite{BL}--\cite{KK} 
to diminish 
the corresponding Born result by a factor of approximately two. 
The reduction 
arises from running the $b$ quark mass from $\overline{m}_b
(m_b)$ up to its 
value $\overline{m}_b (M_H)$ at the Higgs scale .  Here
$\overline{m}_b (\mu)$ 
is the running $b$ quark mass in the $\overline{\rm MS}$ scheme
\cite{BBDM}.

We may write the cross section for the 
background arising from the central production of
quasi-two-jet-like events 
with at least one energetic $b$ jet tagged in the form
\bea
\label{eq:a19}
\sigma \left (\gamma\gamma |_{J_z = 0} \rightarrow 2 \: {\rm
jets} \right 
)_{{\rm single} \: b \: {\rm tag}} & = & \sigma \left
(\gamma\gamma|_{J_z 
= 0} \rightarrow b\overline{b} \right ) \: F_g F_q \; + \;
\sigma \left (
\gamma\gamma |_{J_z = 0} \rightarrow b\overline{b}g \rightarrow 2
\: {\rm jets} 
\right )_{\rm collinear} \nonumber \\
& & \\
& + & \sigma \left ( 
\gamma\gamma|_{J_z = 0} \rightarrow b\overline{b}g \rightarrow 2
\: {\rm jets} 
\right )_{\rm Compton}. \nonumber
\eea
This formula displays the general structure of the background
contributions.\footnote{Note that without a thorough study of the
single logarithmic effects, the question concerning the $b$-quark
mass prescription in the first term in (\ref{eq:a19}) remains
open.} The first term contains a new
non-Sudakov form 
factor $F_q$ which arises from virtual diagrams of the type shown
in Fig.~1.  
The physical origin of this form factor is elucidated in section
4.  In the double logarithmic (DL) approximation $F_q$ has the
form
\be
F_q (L_m) \; = \; \sum_n \: c_n \left ( \frac{\alpha_S}{\pi} \:
L_m^2 
\right )^n
\label{eq:a21}
\ee
with $c_0 = 1$ and $c_1 = -8$ \cite{JT1} so that the second
negative term in 
(\ref{eq:a21}) dominates over the Born term for $M_H \sim 100$
GeV.  This 
dominance undermines the results of analyses
\cite{BBB}--\cite{KMC} which are 
based on the one-loop approximation.  The calculation of the
coefficient $c_2$ 
is one of the aims of the present work, see section 5.  It is
worth noting that 
the same form factor $F_g$ occurs in 
the signal (\ref{eq:a17}) and in the background contribution
(\ref{eq:a19}).  Note from (\ref{eq:b10}) that the non-radiative 
$J_z = 0$ background, the first term in (\ref{eq:a19}), is most 
important for the smaller Higgs masses in the interval given in
(\ref{eq:a1}).

The collinear contribution in (\ref{eq:a19}), in which we have
gluon bremsstrahlung 
off one of the energetic $b$ quarks, can be suppressed by using
the  
$\varepsilon_g$ cut, see Eq.\ (\ref{eq:a12}),(or the traditional
$y_{\rm cut}$)
to 
discriminate between two and three 
jet topologies \cite{BKSO}.  However, we note here, that due to
the form factor 
$F_g$, the imposition of the cut-off will automatically deplete
the signal 
(as well as the first term in (\ref{eq:a19})).  It
could be a non trivial task to find the optimal choice for the
cut-off.

As we have seen in (\ref{eq:a13}) the Compton contribution is
sizeable and should 
be avoided if at all possible by tagging energetic $b$ and
$\overline{b}$ jets.  
Obviously such double tag events have no contribution from the
Compton 
configuration.  The Compton contribution was roughly estimated in
\cite{BKSO} 
by exploiting the fact that it does not have a particularly
strong dependence 
on the helicities of the photons.  Since the Compton regime may
play an 
important role in realistic experiments we calculate the cross
section to 
logarithmic accuracy in the next section. \\

\noindent {\large \bf 3.~~The Compton regime contribution}

As we mentioned above, in the case when only one $b$ jet is
identified the 
$\gamma\gamma \rightarrow b\overline{b}g \rightarrow 2$ jet 
cross section may receive a large contribution from the
configuration where the energetic quark and gluon appear as jets 
in the central detector and the 
spectator quark is comparatively soft and quasi-collinear with
one of the incoming photons 
(the so-called Compton regime \cite{BKSO}).  The size of the
virtual Compton scattering contribution was qualitatively
estimated in
\cite{BKSO} for the case of unpolarized photons.

For polarized photons we use the method of \lq\lq quasi-real
fermions" \cite{VNB} to evaluate the Compton contribution of
Fig.~2 in 
the region where the maximal energy of the unregistered quark
satisfies 
\be
\Delta E_q \; = \; \varepsilon_q \: M_H \; \gg \; m_q.
\label{eq:a22}
\ee
This enables us to write, to logarithmic accuracy, the matrix
element in 
the factorized form 
\bea
\label{eq:a23}
M \biggl (\gamma (\lambda_1, k_1) & + & \gamma (\lambda_2, k_2)
\; \rightarrow 
\; q (h, p) \; + \; \overline{q} (\overline{h}, \overline{p}) \;
+ \; g 
(\lambda_g, k) \biggr )_{\rm Compton \: regime} \nonumber \\
& & \\
& = & \sum_{h^\prime} \: f_{\lambda_1}^{h^\prime \overline{h}}
\: . \: 
\overline{M}_{\rm Compton} \biggl ( \gamma (\lambda_2, k_2) \; +
\; q (h^\prime, k_1 - 
\overline{p}) \; \rightarrow \; q (h, p) \; + \; g
(\lambda_g, k) \biggr ), \nonumber 
\eea
where the helicities ($\lambda$ or $\frac{1}{2} h$) and four
momenta of the 
various particles are indicated in brackets.  The amplitude $f$,
which describes 
the $\gamma (\lambda_1) \rightarrow q (h^\prime) \: \overline{q} 
(\overline{h})$ splitting, is given by
\be
f_{\lambda_1}^{h^\prime \overline{h}} \; = \; - \: \frac{e \:
Q_q}{2 (k_1 . 
\overline{p})} \: \overline{u}^{h^\prime} (k_1 - \overline{p}) \:
\fmslash{e}_{\lambda_1} 
(k_1) \: v^{\overline{h}} (\overline{p}),
\label{eq:a24}
\ee
where $\fmslash{e} \equiv \gamma . e$ and $e^\mu$ is the
polarization vector 
of the photon.  We normalise the four component helicity spinors
so that
\be
(u^h)^\dagger u^h \; = \; 2 p_0, \quad (v^{\overline{h}})^\dagger
\: 
v^{\overline{h}} \; = \; 2
\overline{p}_0.
\label{eq:a25}
\ee
The amplitude $\overline{M}_{\rm Compton}$, which describes the
hard Compton 
subprocess, is evaluated on-mass-shell and all quark masses are
neglected.  
The other three contributions corresponding to the interchanges
$q \leftrightarrow 
\overline{q}$ and/or $k_1 \leftrightarrow k_2$ in Fig.~2 are
obtained from 
(\ref{eq:a23}) and (\ref{eq:a24}) by the obvious substitutions.

We denote the (unnormalized) polarization density matrix of the
incoming 
quark in the Compton subprocess by $\overline{\rho}$.  That is
\be
\overline{\rho}^{h^\prime h^{\prime\prime}} \; = \;
\sum_{\overline{h}} \: 
f^{h^\prime \overline{h}} \left
(f^{h^{\prime\prime} \overline{h}} \right )^*,
\label{eq:a26}
\ee
where for clarity we have omitted the $\lambda_1$ subscript.

Now in the ultrarelativistic limit for small angles
$\overline{\theta}$ of the 
outgoing spectator antiquark with respect to the parent photon
direction 
$\mbox{\boldmath $k$}_1$ we find (\ref{eq:a24}) gives
\bea
\label{eq:a27}
& & \overline{\rho}_{\lambda_1}^{h^\prime h^{\prime\prime}} \;
\simeq \; 
\frac{\pi \alpha Q_q^2}{(k_1 . \overline{p})^2} \: \omega^2 \left
(
\frac{\overline{x}}{1 - \overline{x}} \right ) \nonumber \\
& & \\
& & \;\;\;\;\; \left \{ \left [ \overline{\theta}^2 \left ( (1 -
\overline{x})^2 
\: + \: \overline{x}^2 \: + \: \lambda_1 h^\prime (1 - 2
\overline{x}) \right ) 
\: + \: \frac{m_q^2}{\overline{x}^2} \: (1 + \lambda_1 h^\prime )
\right ] \; 
\delta_{h^\prime, h^{\prime\prime}} \: + \: 2 \lambda_1
\overline{\theta} \: 
\frac{m_q}{\omega} \: \delta_{h^\prime, -h^{\prime\prime}} \right
\}, \nonumber
\eea
where $\overline{x} = \overline{p}_0/\omega$, and $\omega =
\sqrt{s}/2$ is the 
photon energy in the $\gamma\gamma$ c.m.\ frame.  If we note that
the 
polarization of the incoming quark is
$$
\mbox{\boldmath $\zeta$} \; = \; \frac{{\rm Tr} (\mbox{\boldmath
$\sigma$} 
\overline{\rho})}{{\rm Tr} (\overline{\rho})}, 
$$
then the longitudinal component is given by
\be
\zeta_L \; = \; \frac{\overline{\rho}^{1, 1} \: - \:
\overline{\rho}^{\: -1,
-1}}{\overline{\rho}^{1, 1} \: + \: \overline{\rho}^{\: -1, -1}}.
\label{eq:a28}
\ee

We are now ready to turn to the head-on Compton scattering of the
photon $\gamma (\lambda, k_2)$ on the longitudinally polarized
quark $q
(\zeta_L, k_1 - p)$.  In the limit $m_q \rightarrow 0$ it can be
shown (see, 
for example \cite{LT})
\be
\sum_{\lambda_g, h \atop \rm colours} \: \left | \overline{M}
\biggl ( \gamma (\lambda) \: q 
(\zeta_L) \: \rightarrow \: q  (h) \: g (\lambda_g) \biggr )
\right |^2 
\; = \; 2C_F (4 \pi)^2 \: \alpha\alpha_S \: Q_q^2 \left [
\frac{\kappa_2}{\kappa_1} \: 
+ \: \frac{\kappa_1}{\kappa_2} \: + \: \zeta_L \lambda \left (
\frac{\kappa_1}
{\kappa_2} \: - \: \frac{\kappa_2}{\kappa_1} \right ) \right ],
\label{eq:a29}
\ee
where $C_F = (N_c^2 - 1)/2 N_c = \frac{4}{3}$ and
\be
\kappa_1 \; = \; k_2 . (k_1 - \overline{p}), \quad \kappa_2 \; =
\; (k_2 .
p).
\label{eq:a30}
\ee
It is easy to show that
\be
\frac{\kappa_2}{\kappa_1} \; \simeq \; \frac{\omega -
p_0}{\overline{p}_0}
\; = \; \frac{1 - x}{\overline{x}}
\label{eq:a31}
\ee
with $x = p_0/\omega$ and
\be
1 \: + \: \cos \theta \; = \; \frac{2 (1 - x)}{\overline{x}} \:
\frac{(1 - \overline{x})}{x}
\label{eq:a32}
\ee
where $\theta$ is the polar angle between the momenta
$\mbox{\boldmath $k$}_1$ 
of the incoming photon and $\mbox{\boldmath $p$}$ of the outgoing
quark in 
the overall cms.  Note that Eq.\ (\ref{eq:a29}) 
includes the sum over the final and the average over the initial
colour states. 

Using (\ref{eq:a23}), (\ref{eq:a26}) and (\ref{eq:a29}) we find
that the 
contribution to the cross section from the Compton regime shown
in Fig.~2 is
\bea
\label{eq:b32}
\frac{d\sigma_{\rm Compton}}{d \cos \theta} \; = \; \int \:
\frac{d^3
\overline{p}}{(2 \pi)^2 
\: 2 \overline{p}_0} \: \frac{\alpha \alpha_S~C_F (1 -
\overline{x})~Q_q^2}{2 
\omega^2 [2 - \overline{x} (1 - \cos \theta) ]^2} \nonumber \\
& & \nonumber \\
\left [ {\rm Tr} (\overline{\rho}) \: \left (
\frac{\kappa_2}{\kappa_1} \: + \: 
\frac{\kappa_1}{\kappa_2} \right ) \: + \: \lambda~{\rm Tr}
(\sigma_3 
\overline{\rho}) \: \left ( \frac{\kappa_1}{\kappa_2} \: - \:
\frac{\kappa_2}
{\kappa_1} \right ) \right ].
\eea
If we perform the integration over $d^3 \overline{p}$ with the
constraint 
(\ref{eq:a22}) and assume that $\varepsilon_q \ll 1$, then to
logarithmic 
accuracy
\be
\frac{d \sigma_{\rm Compton}}{d \cos \theta} \; = \; \frac{\pi
\alpha^2
Q_q^4}{\omega^2} \: 
\frac{2}{(1 + \cos \theta)} \: \frac{\alpha_S C_F}{\pi} \:
\varepsilon_q~\ln 
\left ( \frac{\varepsilon_q M_H}{m_q} \right ).
\label{eq:c32}
\ee
Summing over all analogous Compton contributions, we obtain for
the final term 
in (\ref{eq:a19}) the explicit expression
\be
\frac{d \sigma_{\rm Compton}}{d \cos \theta} \; = \; \frac{\pi
\alpha^2
Q_q^4}{\omega^2 \sin^2 
\theta} \: \frac{8 \alpha_S C_F}{\pi} \: \varepsilon_q~\ln \left
( 
\frac{\varepsilon_q M_H}{m_q} \right ),
\label{eq:d32}
\ee
where $\alpha_S$ is to be evaluated at the hard scale $M_H$. \\

\noindent {\large \bf 4.~~Physical origin of DL form factors: 
one-loop approximation}

Recall that the $m_q^2$ suppressed term in (\ref{eq:a19})
contains, in addition 
to the standard Sudakov-like form factor, a double logarithmic
form factor $F_q$.  
To leading logarithmic order $F_q$ has the form shown by the
series in 
(\ref{eq:a21}).  The resummation of this DL series looks a very
difficult task.  
In this section we describe the physical origin of the form
factor $F_q$ and 
illustrate the derivation of the coefficient $c_1$.  Then in
section 5 we 
calculate the (positive) $c_2$ coefficient of the series.

The double logarithmic asymptotics of high energy processes has
been the subject 
of intense study even before the birth of QCD\footnote{A detailed
presentation 
of DL results in QED is given, for example, in \cite{LBP,VGG} and
a comprehensive 
QCD review in \cite{QCD}.}.  The most familiar is the so-called
Sudakov form 
factor \cite{VVS} which occurs if we insist on the suppression of
soft collinear 
radiation.  Less frequently we meet other types of DL effects. 
For 
instance, specific DL behaviour appears in high energy $e\mu$
backward 
scattering \cite{GGL},
\be
e (p_1) \: + \: \mu (p_2) \; \rightarrow \; e (p_3) \: + \: \mu
(p_4),
\label{eq:a33}
\ee
as exemplified by one-loop diagram of Fig.~3.  The analogous
process in QCD is 
the backward scattering of two 
quarks of different flavour.  Here soft fermion propagators play
a crucial 
role, whereas the Sudakov form factor arises from soft photon (or
gluon) 
effects.

It is evident that soft virtual fermions can cause DL effects
only in special 
circumstances.  Recall that the boson propagators $D_\gamma, D_g
\sim 1/p^2$ 
while the fermion propagators $D_\ell, D_q \sim 1/\fmslash{p}$ in
the massless 
limit, where $p$ is the particle four momentum.  The special
feature of high 
energy $e\mu$ backward scattering is that the four momentum of
the incoming 
$e (\mu)$ essentially coincides with that of the outgoing $\mu
(e)$, $p_1 \approx 
p_4$ and $p_2 \approx p_3$.  Hence the momenta of the virtual
fermions are 
approximately equal, $p \approx p^\prime$, and so their
propagators double up.  
Then a DL contribution emerges, after the integration over the
soft fermion 
momenta, in a similar way to the emergence of the Sudakov form
factor from the 
integration over the soft boson momentum.  Some interesting
applications of 
these DL effects were originally discussed in \cite{GGL} for QED,
and 
subsequently in \cite{KL,BER} for QCD.

Here we study another manifestation of DL non-Sudakov effects,
namely the form 
factor $F_q$ in the first term of (\ref{eq:a19}).  Thus we are
concerned with 
helicity amplitudes $M_{\lambda, \lambda}^{h, h}$ which contain
an $m_q$ 
suppression factor.  To obtain the DL contribution we therefore
keep $m_q$ in 
one of the fermion propagators which makes the behaviour
$m_q/(p^2 - m_q^2)$ 
similar to that of the boson propagator\footnote{The DL physics
here
resembles the known case (see e.g.\ \cite{GHKD,VVZS}) of the
contribution from the
light fermion loops $(m_f \ll M_H)$ into the $\gamma\gamma$ or
$gg$ partial
widths of a Higgs boson.}.

>From the physical point of view the appearance of a new type of
DL effects in matrix elements with helicity violation is
connected with behaviour of the basic QCD (or QED) transition
amplitudes.  First, helicity-violating $g \rightarrow
q\overline{q}$ or $q \rightarrow qg$ amplitudes do
not vanish when the momenta of the particles become parallel,
unlike the helicity-conserving case, as is clear from angular
momentum conservation.  Secondly, these helicity-violating
amplitudes depend on the energy $\varepsilon$ of the 
softest quark (anti-quark) as $m_q/\sqrt{\varepsilon}$
(supposing, of course, that $\varepsilon \gg m_q$), whereas the
vertices with helicity conservation behave as
$\sqrt{\varepsilon}$.  Therefore, in order to have DL 
effects due to the soft quark it must propagate between two
vertices, one of which is helicity-violating.  Of course, we need
large invariant masses of any pair of other particles entering
the different vertices. Fig.~1 exemplifies such a situation.

In the Feynman gauge the diagram in Fig.~1 also gives a DL
contribution from soft gluon exchange.  However, soft gluon and
soft quark DL effects are very different.  To understand this
difference it is convenient to use a physical gauge.  In this
gauge it is clear that to give DL effects the gluon must connect
two helicity-conserving vertices (because it is soft) and,
moreover, the vertices must occur on the same line (in order to
generate an angular logarithm).  Thus we have a self-energy
diagram.  Therefore the soft gluon DL effects are related to the
real emission of gluons, and we have the well-known cancellation
between virtual and real contributions.

On the other hand, soft quark DL effects in helicity-violating
processes are not related to real emission and so we have no
cancellation.  In summary, soft gluon DL effects have a simple
probabilistic interpretation while no such picture exists in the
case of soft quark DL effects.  Rather the latter are essentially
interference effects.

To be specific, in this section we calculate the one-loop DL
corrections to the 
dominant amplitude, $M_{\lambda\lambda}^{hh}$ with $h = \lambda$,
and leave the 
two-loop effects to section 5. Recall that the amplitude with $h
= -
\lambda$ is suppressed 
by another factor of $m_q/M_H$, see (\ref{eq:a7}).  In fact from
now on we shall only consider the case in which the photon and
quark (antiquark) helicities are all equal to $\lambda$.  We will
therefore omit the helicity subscripts and superscripts from the
amplitudes on the understanding that they are all to be taken
equal to $\lambda$.  Now the box diagram shown in Fig.~1 gives
the ${\cal O} (\alpha_S)$ contributions to the DL form factors
$F_g$ and $F_q$, where $F_g$ is related to the soft gluon 
contribution, while $F_q$ corresponds to the soft virtual quark
contributions.  In general we can evaluate the DL terms by
keeping the dependence on the momentum $k$ of the virtual soft
parton only in its propagator and in the denominators of 
the propagators of the virtual particles joined to the soft
parton.  In the particular case of the ${\cal O} (\alpha_S)$ DL
corrections to $M (\gamma\gamma \rightarrow q\overline{q})$ this
means that we simply have to evaluate the four non-overlapping
kinematic configurations depicted in Fig.~4.  The blobs in these
four diagrams denote the hard $2 \rightarrow 2$ subprocess and
indicate that inside of a blob we can neglect all virtualities
of particles external to the blob.  Note that in Fig.~4(a) helicity
conservation is violated in the hard blob, while in Figs.~4(b-d)
the hard blobs conserve helicity and so we can put $m_b = 0$ when
calculating them.

It is convenient to write the Born amplitude $M_{\rm Born}$ for
central $q\overline{q}$ production in the ultrarelativistic 
limit in a form,
\be
M_{\rm Born} \; = \; - 8e^2 Q_q^2 \: \frac{m_q} {M_H^2} \: 
\frac{\mbox{\boldmath $e$}^\lambda (k_1) \: . \: \mbox{\boldmath
$e$}^\lambda (k_2)}{\sin^2 \theta} \; \overline{u}^\lambda (p) \:
v^\lambda (\overline{p}),
\label{eq:a34}
\ee
which is independent of the phases of the particle wave
functions.

Note that Eq.\ (\ref{eq:a7}) corresponds to the choice of the
polarization vectors $\mbox{\boldmath $e$}^\lambda (k_1)$ and
$\mbox{\boldmath $e$}^\lambda (k_2)$ of the incoming photons in
the $\gamma\gamma$ c.m.\ frame with the $z$ 
axis directed along $\mbox{\boldmath $k$}_1$ defined as in
\cite{LBP}, i.e.\ 
for $\lambda = \lambda_1 = \lambda_2 = \pm 1$
\bea
\label{eq:a35}
\mbox{\boldmath $e$}^\lambda (k_1) & = & - \:
\frac{i\lambda}{\sqrt{2}} \: 
(1, i\lambda, 0), \nonumber \\
& & \\
\mbox{\boldmath $e$}^\lambda (k_2) & = & \mbox{\boldmath $e$}^{-
\: \lambda} 
(k_1). \nonumber
\eea
For such a choice one has
\be
\biggl (\mbox{\boldmath $e$}^\lambda (k_i) \biggr )^* \; = \;
\mbox{\boldmath 
$e$}^{- \: \lambda} (k_i); \quad\quad \fmslash{\mbox{\boldmath
$e$}}^\lambda 
(k_2) \: \fmslash{\mbox{\boldmath $e$}}^{- \: \lambda} (k_1) \; =
\; 0,
\label{eq:b35}
\ee
where $\fmslash{\mbox{\boldmath $e$}} \equiv \mbox{\boldmath
$\gamma$} . 
\mbox{\boldmath $e$}$.  We define the quark and antiquark 
wavefunctions as
\bea
\label{eq:c35}
u^\lambda (p) & = & \sqrt{p_0 \: + \: m_q} \; \left
(\begin{array}{c}
\phi_\lambda (\mbox{\boldmath $p$}) \\  \lambda |\mbox{\boldmath
$p$}| \phi_\lambda (\mbox{\boldmath $p$})/(p_0 + m_q) \end{array}
\right ), \nonumber \\
& & \nonumber \\
v^\lambda (\overline{p}) & = & \sqrt{\overline{p}_0 + m_q} \;
\left (\begin{array}{c}
-\lambda |\mbox{\boldmath $\overline{p}$}| \phi_\lambda
(-\mbox{\boldmath $\overline{p}$})/(\overline{p}_0 + m_q) 
\\ \phi_\lambda (- \: \mbox{\boldmath $\overline{p}$})
\end{array} \right )
\eea
where, as usual, we take $\lambda$ to be the (double) quark
helicity, that is
\be
\frac{\mbox{\boldmath $\sigma$} . \mbox{\boldmath
$p$}}{|\mbox{\boldmath $p$}|} \: \phi_\lambda (\mbox{\boldmath
$p$}) \; = \; \lambda \: \phi_\lambda (\mbox{\boldmath $p$}).
\label{eq:d35}
\ee
For virtual gluons we use the Feynman gauge.

The amplitudes corresponding to each of the four diagrams $(i =
a,b,c,d)$ in Fig.~4 can be written in the factorized form
\be
M_i \; = \; {\cal F}_i \: M_{\rm Born},
\label{eq:a36}
\ee
where ${\cal F}_i$ is the form factor from diagram $i$.  We
elucidate this result below, taking the diagrams in turn.  We
stress that each amplitude $M_i$ (apart from $M_a$) describes the
sum of the
corresponding diagram in Fig.~4 and the crossed diagram with the
photon momenta interchanged, $k_1 \leftrightarrow k_2$.

\medskip
\noindent {\bf 4.1~~The form factor with the $\gamma\gamma
\rightarrow q\overline{q}$ hard subprocess}

The Sudakov DL effects which arise from the virtual soft gluon
are shown in Fig.~4(a).  Here, for $M_a$, the factorized form
(\ref{eq:a36}) is immediately evident.  In this case
\be
{\cal F}_a \; = \; 4 \pi \alpha_S C_F \: \int \: \frac{d^4
k}{i (2 \pi)^4} \; \frac{-4 \: (p . \overline{p})}{[k^2 -
m_g^2 + i \varepsilon][(p + k)^2 - m_q^2 + i
\varepsilon][(\overline{p} - k)^2 - m_q^2 + i\varepsilon]},
\label{eq:a37}
\ee
where $C_F = \frac{4}{3}$ and where a gluon mass $m_g$ is
introduced to regularize the infrared singularity.  Then, on
carrying out the integration in the standard way, we obtain the
well known DL result
\be
{\cal F}_a \; = \; - \frac{\alpha_S}{\pi} \: C_F \left (
L_m^2 \: + \: 
L_m \ln \frac{m_q^2}{m_g^2} \right )
\label{eq:a38}
\ee
where we recall that $L_m \equiv \ln (M_H/m_q)$.  As usual the
infrared divergence $\ln m_g^2$, and also the $\ln^2 m_q^2$
term, cancels after adding the soft real gluon emission
contribution.  We return to discuss this cancellation at the end
of this section.

At first sight the factorized form (\ref{eq:a36}) is not so
obvious for diagrams 4(b,c,d) and so we derive it below.  Recall
that the overall $m_q$ suppression of these amplitudes comes from
the fact that only the mass term in the numerator of the
propagator $D_q (k) \simeq (\fmslash{k} + m_q)/(k^2 - m_q^2)$ 
of the soft quark can contribute and we need retain $k$ only in
the denominators of the propagators of the virtual particles
joined to the
soft quark.

\medskip
\noindent {\bf 4.2~~The form factor with the $q\overline{q}
\rightarrow q\overline{q}$ hard subprocess}

First we study the amplitude corresponding to Fig.~4(b).  We may
neglect the mass in the numerators of propagators $D_q (k_i)$ of
the quarks joined to the soft quark with $i = 1, 2$ and write
them as
\be
\fmslash{k}_i \; = \; \sum_\lambda \: u^\lambda (k_i) \:
\overline{u}^\lambda (k_i) \; = \; \sum_\lambda \: v^\lambda
(k_i) \:\overline{v}^\lambda (k_i),
\label{eq:a40}
\ee
and use the relations
\be
\label{eq:a39}
\mbox{\boldmath $\fmslash{e}$}^\lambda (k_i) \: v^{- \lambda}
(k_i) \; = \; 
\mbox{\boldmath $\fmslash{e}$}^\lambda (k_i) \: u^\lambda (k_i)
\; = \; 
\overline{v}^\lambda (k_i) \: \mbox{\boldmath
$\fmslash{e}$}^\lambda (k_i) \; 
= \; \overline{u}^{\: -\lambda} (k_i) \: \mbox{\boldmath
$\fmslash{e}$}^\lambda 
(k_i) \; = \; 0 
\ee
in the massless limit.  Then it can be easily seen that the 
amplitude for diagram 4b contains the $q\overline{q} \rightarrow
q\overline{q}$ amplitude as a factor.  To be precise we have
\bea
\label{eq:a41}
M_b & \simeq & \left ( \frac{\alpha}{\alpha_S} \right ) \: Q_q^2
\: {\cal F}_b \: M \biggr (q (k_1, \lambda) + \overline{q} (k_2,
\lambda) \: \rightarrow \: q (p, \lambda) + \overline{q}
(\overline{p}, \lambda) \biggr ) \nonumber \\
& & \\
& & \times \; \frac{\overline{u}^\lambda (k_1) \:
\fmslash{e}^\lambda (k_1) \: m_q \: \fmslash{e}^\lambda (k_2) \:
v^\lambda (k_2)}{4 C_F (k_1 . k_2)} \;\; + \;\; \biggl \{ \: k_1
\leftrightarrow k_2 \: \biggr \}. \nonumber
\eea
The summation over the colours of the intermediate $q (k_1,
\lambda)$ and $\overline{q} (k_2, \lambda)$ states is implied
here.  The effect is to cancel the factor $C_F$ in the
denominator.  This enables us to have the same normalisation of
the form factors ${\cal F}_i$ of diagrams 4(a,b,c).  Due to the
conservation of colour only the scattering diagram (and not the
annihilation diagram) contributes to $M (q\overline{q}
\rightarrow q\overline{q})$ in (\ref{eq:a41}).  Note that the 
annihilation channel is also suppressed due to helicity
conservation.

The DL factor in (\ref{eq:a41}) is given by
\bea
\label{eq:a42}
{\cal F}_b & = & 4\pi \alpha_S (M_H) C_F \: \int \:
\frac{d^4 k}{i (2\pi)^4} \; \frac{-4 (k_1 . k_2)}{[k^2 - m_q^2 +
i\varepsilon][(k_1 + k)^2 - m_q^2 + i\varepsilon][(k_2 - k)^2 -
m_q^2 + i\varepsilon]} \nonumber \\
& & \nonumber \\
& = & - \: \frac{\alpha_S (M_H)}{\pi} \: C_F L_m^2 \equiv {\cal
F}.
\eea

To reduce (\ref{eq:a41}) to the factorized form given in
(\ref{eq:a36}) we rewrite the $M (q\overline{q} \rightarrow
q\overline{q})$ scattering amplitude using the Fierz
transformation (which is especially simple when the particles 
have the same helicities)
\be
(\overline{u}_1^\lambda \: \gamma^\mu \:
u_2^\lambda)(\overline{v}_3^\lambda \: 
\gamma_\mu \: v_4^\lambda) \; = \; 2 (\overline{u}_1^\lambda \:
v_4^\lambda)
(\overline{v}_3^\lambda \: u_2^\lambda).
\label{eq:a43}
\ee
Then the numerator in (\ref{eq:a41}) can be rearranged to contain
the factor
\be
\label{eq:a44}
{\rm Tr} \biggl (\fmslash{k}_1 \: \fmslash{e}^\lambda (k_1) \:
\fmslash{e}^\lambda 
(k_2) \: \fmslash{k}_2 (1 + \lambda \gamma_5) \biggr ) \:
\overline{u}^\lambda 
(p) \: v^\lambda (\overline{p}) \; = \; -8 (k_1 . k_2) \: \biggl
( \mbox{\boldmath $e$}^\lambda (k_1) \: . \: \mbox{\boldmath
$e$}^\lambda (k_2) \biggr ) \: \overline{u}^\lambda (p) \:
v^\lambda (\overline{p}).
\ee
If we use this result, together with (\ref{eq:a34}), we find
(\ref{eq:a41}) has 
the factorized form given in (\ref{eq:a36}) for $i = b$.

\medskip
\noindent {\bf 4.3~~The form factors for the Compton scattering
hard subprocesses}

We now turn to the final two diagrams of Fig.~4.  Noting the
conservation of quark helicity in the hard process, we obtain
\bea
\label{eq:a45}
M_c & = & \sqrt{\frac{\alpha}{\alpha_S}} \: Q_q \: {\cal F}_c \:
M^\mu \biggl ( \gamma (k_1, \lambda) \: + \: q (k_2, \lambda) \:
\rightarrow \: q (p, \lambda) \: + \: g (\overline{p}, \mu) 
\biggr ) \nonumber \\
& & \\
& \times & \frac{\overline{u}^\lambda (k_2) \:
\fmslash{e}^\lambda (k_2) \: m_q \: \gamma_\mu \: v^\lambda
(\overline{p})}{4 C_F (\overline{p} . k_2)} \;\; + \;\; \biggl \{
\: k_1 \leftrightarrow k_2 \: \biggr \}. \nonumber
\eea
where $M^\mu$ is the Compton $(\gamma q \rightarrow qg)$
amplitude for a gluon 
with polarization index $\mu$.  Once again a summation of the
colours of the intermediate particles (the quark and the gluon)
is implied.  Again it leads to a cancellation of the colour
factor $C_F$ shown in (\ref{eq:a45}).  The form factor ${\cal
F}_c$ is given by
\be
{\cal F}_c \; = \; 4 \pi \alpha_S C_F \: \int \: \frac{d^4
k}{i (2\pi)^4} 
\; \frac{4 (k_2 . \overline{p})}{[k^2 - m_q^2 +
i\varepsilon][(k_2 + 
k)^2 - m_q^2 + i\varepsilon][(\overline{p} + k)^2 +
i\varepsilon]}.
\label{eq:a46}
\ee
It is useful to note that in the one-loop approximation only the
$s$-channel 
diagram contributes to the amplitude for the Compton scattering. 
The $u$-channel 
contribution vanishes in this approximation due to relation
\be
\label{eq:b46}
2 \gamma^\mu \: u^\lambda (k_2) \: \overline{u}^\lambda (k_2) \: 
\fmslash{e}^\lambda (k_2) \: \gamma_\mu \; = \; \gamma^\mu \:
\fmslash{k}_2 
(1 - \lambda \: \gamma_5) \: \fmslash{e}^\lambda (k_2) \:
\gamma_\mu \; = \; 
4 (1 - \lambda \: \gamma_5) (k_2 . e^\lambda (k_2)) \; = \; 0.
\ee
However, in higher order one has to take into account both the
$s$ and $u$ 
channel contributions.

It is easy to see that both diagram 4c and the crossed $(k_1 
\leftrightarrow k_2)$ diagram contain DL factors which, to DL
accuracy, are 
equal to each other.  Moreover the integrands in (\ref{eq:a42})
and (\ref{eq:a46}) 
are related by the interchange
\be
k_1 \; \leftrightarrow - \overline{p}, \quad\quad k \;
\rightarrow \; - k.
\label{eq:a47}
\ee
To the accuracy to which we are working, the absence of $m_q^2$
in the third 
factor of the denominator in (\ref{eq:a46}) is not important. 
Thus we have
\be
{\cal F}_c \; = \; - \frac{\alpha_S}{\pi} \: C_F \: L_m^2 = {\cal
F}.
\label{eq:a48}
\ee
Note that for ${\cal F}_c$ (and also ${\cal F}_{a,d}$) we have no
reason to 
evaluate $\alpha_S$ at the hard scale $M_H$, although in the
following we shall 
not take into account possible differences in the scale of
$\alpha_S$ for the 
different ${\cal F}_i$.  If we manipulate the spinor
structure in (\ref{eq:a45}), just as we did for the previous case
of amplitude $M_b$, then we obtain the factorized form
(\ref{eq:a36}) for $M_c$ also.
By repeating the same procedure, it is straightforward to
show that the factorized form (\ref{eq:a36}) follows for the
amplitude $M_d$ as well and that
\be
{\cal F}_d \; = \; {\cal F}_c = {\cal F}.
\label{eq:a49}
\ee

It is worth drawing attention to one subtlety.  We had noted that
the amplitude $M_b$ contained as one of the factors the physical
$q\overline{q} \rightarrow q\overline{q}$ amplitude with the same
helicities for the incoming and outgoing quarks, see
(\ref{eq:a41}).  The same is not true for $M_c$.  In
(\ref{eq:a45}) we sum over all the gluon polarization states,
including the non-physical ones.  However, it is possible to
restore the symmetry of the helicity structure of $M_c$ to that
of $M_b$ if we exploit the fact that both $M^\mu (\gamma q
\rightarrow qg)$ and its spinor multiplier in (\ref{eq:a45})
vanish if multiplied by $\overline{p}_\mu$.  The former follows
from gauge invariance, $\overline{p}_\mu M^\mu = 0$ and the
latter is a consequence of the Dirac equation for $v^\lambda 
(\overline{p})$.  Thus we need only sum over the physical gluon
polarization states $e_\mu^\lambda (\overline{p})$,
which satisfy
\be
\fmslash{\varepsilon}^{- \lambda} (\overline{p}) \: v^\lambda
(\overline{p}) \; = \; 0.
\label{eq:a53}
\ee
This condition is gauge invariant and it can be derived in the
same way as (\ref{eq:a39}).  Using these results we can rearrange
(\ref{eq:a45}) into the form
\bea
\label{eq:a54}
M_c & = & -~\sqrt{\frac{\alpha}{\alpha_S}} \: Q_q \: {\cal F}_c
\: M \biggl ( \gamma (k_1, \lambda) + q (k_2, \lambda) \:
\rightarrow \: q (p, \lambda) + g (\overline{p}, \lambda) \biggr
) \nonumber \\
& & \\
& \times & \frac{\overline{u}^\lambda (k_2) \:
\fmslash{e}^\lambda (k_2) \: m_q \: \fmslash{e}^\lambda
(\overline{p}) \: v^\lambda (\overline{p})}{4 C_F (\overline{p} .
k_2)} \;\; + \;\; \biggl \{ \: k_1 \leftrightarrow k_2 \: \biggr
\} \nonumber
\eea
in which we keep only the contribution corresponding to a gluon
of helicity $\lambda$.

Before proceeding to the study of higher loop contributions in
section 5, note that although the derivation of the factorized
form (\ref{eq:a36}) for the amplitudes $M_{b,c,d}$ was given in a
way that allowed a clear physical interpretation, it is probable
that there is a more general reason for this result.  It could
well be that factorized relations of the type shown in
(\ref{eq:a41}) and (\ref{eq:a54}) are the result of
(super)symmetry relations between amplitudes with the same 
values of the helicities (and doubled helicities) of the bosons
(and fermions) participating in the hard scattering.

\medskip
\noindent {\bf 4.4~~Expressions for the form factors $F_g$ and
$F_q$}

We return to the calculation of the one-loop correction to the
cross section.  On account of the factorized form (\ref{eq:a36}),
the one-loop virtual corrections give a factor $(1 + \delta_V)$
in the formula for the cross section, where
\be
\delta_V \; = \; 2 \: \sum_{i = a,b,c,d} \: {\cal F}_i.
\label{eq:i1}
\ee
We must combine this correction with the contribution from soft
real gluon emission.  Now the amplitude $M^{(1)}$ describing the
emission of {\it one} soft gluon of momentum $k_g$ is
\be
M^{(1)} \; = \; M_{\rm Born} \: g_S \langle t^a \rangle \: e_\mu
(k_g) \: J^\mu (k_g)
\label{eq:i2}
\ee
with
\be
J^\mu (k) \; = \; \frac{p^\mu}{p . k} \: - \:
\frac{\overline{p}^\mu}{\overline{p} . k},
\label{eq:i3}
\ee
where $e_\mu$ is the polarisation vector of the emitted
gluon; and $p, \overline{p}$ are the momenta of outgoing quark
and antiquark.  The gluon has colour index $a$, and $\langle t^a
\rangle$ denotes the generator of the fundamental representation
of the colour group evaluated between the $q$ and $\overline{q}$
states.  Therefore the correction to the cross section due to
real emission is
\bea
\label{eq:i4}
\delta_R & = & g_S^2 \: C_F \: \int_{\Omega_R} \: \frac{d^3 k}{(2
\pi)^3 2\omega} \: \left ( - J_\mu (k) \: J^\mu (k) \right )
\nonumber \\
& & \nonumber \\
& = & \frac{\alpha_S C_F}{4 \pi^2} \: \int_{\Omega_R} \:
\frac{d^3 k}{\omega} \: \frac{2 (p . \overline{p})}{(k . p) (k .
\overline{p})},
\eea
with $\omega = (\mbox{\boldmath $k$}^2 +
m_g^2)^{\frac{1}{2}}$, and where $\Omega_R$ denotes the
region of phase space of gluon radiation allowed by the cut-off
prescription.

The infrared divergence in $\delta_R$ cancels that of the virtual
contribution $2 {\cal F}_a$ in (\ref{eq:i1}).  Indeed the sum
$\delta_R + 2 {\cal F}_a$ represents the first term of the
expansion of the Sudakov form factor $F_g$ which occurs in
(\ref{eq:a17}) and (\ref{eq:a19}).  Hence we have
\be
F_g \; = \; \exp (\delta_R + 2 {\cal F}_a).
\label{eq:i5}
\ee
For the non-Sudakov form factor $F_q$ we have, in the one-loop
approximation,
\be
F_q \; = \; 1 \: + \: 2~\sum_{i = b,c,d} \: {\cal F}_i
\; = \; 1 \: + \: 6 {\cal F}
\; = \; 1
\: - \: \frac{8 \alpha_S}{\pi} \: L_m^2.
\label{eq:i6}
\ee
where we have used (\ref{eq:a42}), (\ref{eq:a48}) and
(\ref{eq:a49}).  That is we have reproduced the result that was
first derived in Ref.\ \cite{JT1}.

Let us now give the explicit expressions for the Sudakov form
factor
$F_g$ for the different cut-off prescriptions. When we impose the
$\varepsilon_g$ restriction
(\ref{eq:a11}) on the gluon transverse energy we obtain
\be
\delta_R (\varepsilon_g) \; = \; - 2 {\cal F}_a \: - \:
2~\frac{\alpha_S C_F}{\pi} \: L_g^2
\label{eq:e54}
\ee
for $|k_{g \perp} | = \Delta E_g \gapproxeq m_q$, where
logarithmic
factor $L_g$ is defined in (\ref{eq:b17}).  Thus on combining
$\delta_R$ with $\delta_V$ of (\ref{eq:i1}) we see that
(\ref{eq:i5}) becomes
\be
F_g (\varepsilon_g) \; = \; \exp \left ( -~\frac{2 \alpha_S}{\pi}
\: C_F \: L_g^2 \right ).
\label{eq:f54}
\ee
On the other hand if we restrict the gluon momentum by $k_g \leq
k_0$, then
\be
\delta_R (k_0) \; = \; - 2 {\cal F}_a \: - \: 4~\frac{\alpha_S
C_F}{\pi} \: L_m L_0,
\label{eq:g54}
\ee
where now the logarithmic factors are defined in (\ref{eq:a18}). 
In this case we see that (\ref{eq:i5}) is of the form
\be
F_g (k_0) \; = \; \exp \left ( -~\frac{4 \alpha_S}{\pi} \: C_F \:
L_m L_0 \right ).
\label{eq:h54}
\ee

Finally, let us present the expression for the Sudakov form
factor in terms of the standard jet-finding parameter $y_{\rm
cut}$ used in Refs.\ \cite{BKSO} - \cite{KMC}. By imposing the
constraint
\be
(p(\bar{p}) + k_g)^2 \: < \: y_{\rm cut} \: s
\label{eq:h55}
\ee
\noindent on the process $\gamma \gamma \: \rightarrow \: q
\bar{q} g$ at the partonic level we can write down the one-loop
real correction $\delta_R$ as
\be
\delta_R (y_{\rm cut}) \; = \; -2 {\cal F}_a \: - \: 
\frac{\alpha_S C_F}{\pi} \; {\rm ln}^2 \; 
\frac{1}{y_{\rm cut}}.
\label{eq:h56}
\ee
\noindent Then (\ref{eq:i5}) becomes
\be
F_g (y_{\rm cut}) \; = \; \exp \left ( -~\frac{\alpha_S}{\pi}\;
C_F \; {\rm ln}^2 \; 
\frac{1}{y_{\rm cut}} \right ).
\label{eq:h57}
\ee
\noindent where we have assumed that $y_{\rm cut} \gg
{m_q^2}/s$. Recall that in the case of form factor $F_g$
there is no reason to evaluate $\alpha_S$ at the hard scale
$M_H$.

\bigskip
\noindent {\large \bf 5.~~The two-loop approximation for
$\gamma\gamma \rightarrow q\overline{q}$}

In this section we study the two-loop approximation to the cross
section for the central production of a $q\overline{q}$ pair in
the collision of photons with equal helicities.  Of course,
$q\overline{q}$ production can be accompanied by other final
state particles, depending on the experimental conditions.  Here
we adopt restrictive experimental criteria so as to provide
conditions that are most favourable for the detection of the
Higgs boson.  This means that, besides the $q\overline{q}$ pair,
the final state contains only soft gluons.  Therefore the
two-loop contribution to the $J_z = 0$ cross section that we are
interested in may be written in the form
\be
d \sigma_{\rm 2-loop} \; = \; d \sigma_{\rm Born} (\gamma\gamma
\rightarrow q\overline{q} gg) \: + \: d\sigma_{\rm 1-loop}
(\gamma\gamma \rightarrow q\overline{q}g) \: + \: d \sigma_{\rm
2-loop} (\gamma\gamma \rightarrow q\overline{q}),
\label{eq:e1}
\ee
where the emitted gluons are soft.  We evaluate the three
components in turn.

\medskip
\noindent {\bf 5.1~~The contribution from $\gamma\gamma
\rightarrow q\overline{q} gg$}

There are two types of DL contribution to the two-gluon emission
component of the cross section of (\ref{eq:e1}) in the double
logarithm approximation.  The first part, part A, comes when the
emitted gluons are strongly ordered in angle.  It is similar to
the QED case and has amplitude
\be
M_A^{(2)} \; = \; M_{\rm Born} \: g_S^2 \: e_\mu (k_{1g}) \:
J^\mu
(k_{1g}) \: e_\mu (k_{2g}) \: J^\mu (k_{2g}) \: C (a_1,
a_2),
\label{eq:e2}
\ee
where the superscript (2) denotes the emission of {\it two} soft
gluons and where $J^\mu$ is defined in (\ref{eq:i3}).  The colour
factor
\be
C (a_1, a_2) \; = \; \left \{ \begin{array}{lcl}
\langle t^{a_1} \: t^{a_2} \rangle & \hbox{if} & \theta_1 \: \ll
\: \theta_2 \\
\langle t^{a_2} \: t^{a_1} \rangle & \hbox{if} & \theta_2 \: \ll
\: \theta_1, \end{array} \right .
\label{eq:e2a}
\ee
where $a_i$ are the colour labels of the emitted gluons and
$\theta_i$ is the angle of the $i$-th gluon with
respect to the quark momentum $\mbox{\boldmath $p$}$.  In both
cases when $|C|^2$ is summed over the colours of the emitted
gluons we obtain $C_F^2$.  Thus part A of the cross section is
\be
d \sigma_{\rm Born}^A (\gamma\gamma \rightarrow q\overline{q} gg)
\; = \; d \sigma_{\rm Born} (J_z = 0) \: \frac{\delta_R^2}{2},
\label{eq:e3}
\ee
where $\delta_R$ is given by (\ref{eq:i4}).

However, unlike QED, there is another region which leads to a
double logarithmic contribution to the cross section.  Namely the
region in which the emission of the two gluons is strongly
correlated so that the angle between their momenta is much less
than the angles of their emission with respect to the quark or
antiquark.  The amplitude in this case (part B) is
\be
M_B^{(2)} \; = \; M_{\rm Born} \: g_S^2 \: \frac{e_\mu (k_{1g})
\: J^\mu (k_{1g}) \: e_\mu (k_{2g}) \: k_{1g}^\mu}{(k_{1g}
. k_{2g})} \; i f_{a_1 a_2 a} \langle t^a \rangle,
\label{eq:e4}
\ee
where $f_{abc}$ is the usual QCD structure constant, and where we
have assumed that the gluon energies satisfy $\omega_{1g} \gg
\omega_{2g}$.  Note that physical polarisation vectors have been
used for the second gluon in (\ref{eq:e4}) which satisfy
$e (k_{2g}) . k_{2g} = 0$.  The contribution to the
cross section from this region, region B, with two soft gluons
emitted is
\bea
\label{eq:e5}
d \sigma_{\rm Born}^B (\gamma\gamma \rightarrow q\overline{q} gg)
& = & d \sigma_{\rm Born} (J_z = 0) \: \frac{\alpha_S C_F}{4
\pi^2} \nonumber \\
& & \nonumber \\
& \times & \int_{\Omega_R} \: \frac{d^3 k}{\omega} \: \frac{2 (p
. \overline{p})}{(k . p) (k . \overline{p})} \: \frac{\alpha_S
C_A}{4 \pi} \: \ln^2 \left ( \frac{(k . p) (k .
\overline{p})}{m_g^2 (p . \overline{p})} \right ),
\eea
where $C_A = N_c = 3$.  The total $\gamma\gamma \rightarrow
q\overline{q} gg$ cross section is the sum of (\ref{eq:e3}) and
(\ref{eq:e5}),
\be
d \sigma_{\rm Born} (\gamma\gamma \rightarrow q\overline{q} gg)
\; = \; d \sigma_{\rm Born}^A \: + \: d \sigma_{\rm Born}^B.
\label{eq:e6}
\ee

\medskip
\noindent {\bf 5.2~~The two-loop contribution from $\gamma\gamma
\rightarrow q\overline{q}g$}

The component of the two-loop cross section (\ref{eq:e1}) which
arises from the emission of a single soft gluon comes from the
interference of $M_{\rm 1-loop}^{(1)}$ with its Born value
$M^{(1)}$ given in (\ref{eq:i2}).  Recall that the superscript
(1) denotes the emission of a {\it single} soft gluon.  Thus we
have to calculate the one-loop correction of the Born amplitude
$M^{(1)}$ for the process $\gamma\gamma \rightarrow
q\overline{q}g$.  In the Born approximation the double
logarithmic contribution to the cross section comes from regions
where the soft gluon is emitted quasi-collinearly with either the
outgoing quark or antiquark.  Because the contributions for
emission along the quark or antiquark direction are equal, we
could restrict ourselves to considering the region of
quasi-collinearity with the quark.  However, we shall not do this
in order to obtain a general picture and to maintain gauge
invariance.

Just as we did in the calculation of the one-loop correction to
the matrix element of the basic $\gamma\gamma \rightarrow
q\overline{q}$ process, we separate the \lq\lq hard" stage of the
process from the \lq\lq soft" stage.  The later stage leads to
double logarithmic corrections.  The gluon is emitted in the soft
stage.  In the cases when the \lq\lq hard" subprocesses do not
coincide with the basic process (namely for Figs.~4(b-d)) the
appropriate diagrams are obtained by the addition of a gluon line
to all the diagrams of Figs.~4(b-d), noting that it cannot be
emitted from the \lq\lq hard blob".  But such a statement is not
correct for the case of the \lq\lq hard" $\gamma\gamma
\rightarrow q\overline{q}$ process of Fig.~4(a).  The reason for
this difference is evident.  In the latter case, contrary to the
former ones, there are other one-loop diagrams besides those
shown in Fig.~4 (for example, diagrams with self-energy
insertions).  We did not consider them because they do not give
DL contributions (in the Feynman gauge which we use here). 
But after the addition of the real soft gluon line they can give
such a contribution.

In fact, the case of the \lq\lq hard" $\gamma\gamma \rightarrow
q\overline{q}$ process of Fig.~4(a) is analogous to the decay of
a \lq\lq heavy" photon into a $q\overline{q}$ pair \cite{KF}. 
The contributing diagrams are shown in Fig.~5.  We also have
diagrams 5($\overline{\rm a},\overline{\rm b},\overline{\rm d}$)
in which the soft gluon is
emitted from the $\overline{q}$
rather than the $q$.  The evaluation of these seven diagrams is
similar to that performed in \cite{KF}.  The result is
\be
M_{\rm 1-loop}^{(1)~{\rm Fig.5}} \; = \; M_{\rm Born} \: g_S
\langle t^a \rangle \: e_\mu (k_g) \: J^\mu (k_g) \: \left [{\cal
F}_a \: - \: \frac{\alpha_S}{2 \pi} \: \frac{C_A}{4} \: \ln^2
\left ( \frac{(k_g . p)(k_g . \overline{p})}{m_g^2 (p .
\overline{p})} \right ) \right ],
\label{eq:e7}
\ee
\noindent where ${\cal F}_a$ is given by (\ref{eq:a38}), $J^\mu$
by
(\ref{eq:i3}) and $m_g$ is the \lq mass' of the gluon. 
Note that the second term in the square brackets violates the
soft emission factorization and Poisson distribution theorems
that hold for QED \cite{KF}.

Next we consider the gluon emission in the case of the
\lq\lq hard"
$q\overline{q} \rightarrow q\overline{q}$ process of Fig.~4(b). 
The diagrams are displayed in Fig.~6.  Again we must also include
the contribution of diagram 6($\overline{\rm a}$) in which the
gluon is emitted from the $\overline{q}$.  The evaluation of the
diagrams, together with the crossed diagrams with $k_1
\leftrightarrow k_2$, is non-trivial and introduces novel
features.  The derivation is described in Appendix A.  The
final result can be presented in the form of Eq.\ (\ref{eq:a41})
with the replacements
\bea
\label{eq:e8}
M_b & \rightarrow & M_{\rm 1-loop}^{\rm (1) Fig.6}, \nonumber \\
& & \nonumber \\
{\cal F}_b & \rightarrow & g_S \langle t^a \rangle \: \biggl [
e_\mu (k_g) \: J^\mu (k_g) \: {\cal F}_b \biggr . \nonumber \\
& & \nonumber \\
& & + \: \biggl . i \lambda \: \frac{\varepsilon_{\mu \nu \rho
\sigma} \: k_2^\mu \: k_1^\nu \: e^\rho (k_g) \: k_g^\sigma}{(k_g
. k_1) (k_g . k_2)} \left ( \frac{- \alpha_S}{4 \pi} \right )
\left (C_F - \frac{C_A}{2} \right ) \ln^2 \left ( \frac{(k_g .
k_1) (k_g . k_2)}{m_q^2 \: (k_1 . k_2)} \right ) \biggr ],
\eea
where ${\cal F}_b$ on the right hand side is given by
(\ref{eq:a42}), $\varepsilon_{\mu \nu \rho \sigma}$ is the
totally antisymmetric tensor, $\varepsilon_{0123} = 1$, and
$\lambda$ specifies the helicity of all the
$\gamma\gamma \rightarrow q\overline{q}$ particles.  The first
term in the square brackets is standard.  It arises from diagrams
6(a, $\overline{\rm a}$).  It is worth noting that the
$\varepsilon_{\mu \nu \rho \sigma}$ term, which contains
correlations involving the photon helicities, are coloured
suppressed by ${\cal O} (1/N_c^2)$ compared to the standard
QED-like terms.  Let us emphasise that this term should be
retained only when $(k_g . k_1)(k_g . k_2) \gg m_q^2 (k_1 .
k_2)$.  It is also assumed here and in what follows that each
term containing a logarithmic factor should be retained only when
the argument of this logarithm is large.

Figs.~6(b,c,d), and their $k_1 \leftrightarrow k_2$ counterparts,
have contributions proportional to
\be
e (k_g) . k_i/(k_g . k_i)
\label{eq:e9}
\ee
with $i = 1,2$, which cancel when we take the sum.  The
cancellation could be naively expected since collinear
photon-gluon
singularities are clearly unphysical.  However, diagram 6(d) also
gives rise to the final term in (\ref{eq:e8}).  This is a novel
contribution.  Evidently its appearance is connected with the
peculiarity of processes with helicity violation.  Due to the
presence of $\varepsilon_{\mu \nu \rho \sigma}$ this contribution
is antisymmetric with respect to the interchange $k_1
\leftrightarrow k_2$.  However, symmetry of ${\cal F}_b$ was
necessary to obtain the factorized form (\ref{eq:a36}) from
(\ref{eq:a41}).  As a consequence diagrams 6(b,c,d) considered
together with the crossed diagrams ($k_1 \leftrightarrow k_2$) do
not have the factorized form shown in (\ref{eq:a36}).

We now turn to the corrections to diagram 4(c) with the \lq\lq
hard" Compton subprocess $\gamma q \rightarrow qg$.  In this case
the resulting diagrams for the gluon emission are shown in
Fig.~7.  The structure of the
final result from the sum of these diagrams is similar to that
obtained from the diagrams of Fig.~6.  It is derived in Appendix
A and corresponds to making the replacements
\bea
\label{eq:e10}
M_c & \rightarrow & M_{\rm 1-loop}^{\rm (1) Fig.7}, \nonumber \\
& & \nonumber \\
{\cal F}_c & \rightarrow & g_S \langle t^a \rangle \: \biggl [
e_\mu (k_g) \: J^\mu (k_g) \: {\cal F}_c \biggr . \nonumber \\
& & \nonumber \\
& & + \: \biggl . i\lambda \: \frac{\varepsilon_{\mu \nu \rho
\sigma} \: \overline{p}^\mu \: k_2^\nu \: e^\rho (k_g) \:
k_g^\sigma}{(k_g . k_2) (k_g . \overline{p})} \left ( \frac{-
\alpha_S}{4 \pi} \right ) \left (C_F - \frac{C_A}{2} \right )
\ln^2 \left ( \frac{(k_g . \overline{p})(k_g . k_2)}{m_q^2 \:
(k_2 . \overline{p})} \right ) \biggr ]
\eea
in (\ref{eq:a45}).  Here $\lambda$ is the helicity, $J^\mu$ is
defined by (\ref{eq:i3}) and ${\cal F}_c$ on the right hand side
is given by (\ref{eq:a48}).  Now the first term comes not only
from the diagrams shown in Figs.~7(a,b), but includes
contributions proportional to $(e_\mu 
(k_g). \overline{p})/(k_g. \overline{p})$ coming from the diagrams 
of Figs.~7(d,e) as well.  The contributions of the type (\ref{eq:e9}) from 
the diagrams of Figs.~7(c,e) 
and their $k_1 \leftrightarrow k_2$ counterparts cancel just as in
the previous case.  The final term comes from the diagram of
Fig.~7(e).

Finally we consider the soft gluon emission in the \lq\lq hard"
Compton subprocess $\gamma \overline{q} \rightarrow
\overline{q}g$ of Fig.~4(d).  This contribution is exactly
analogous to the previous one and can be obtained by making in
(\ref{eq:e10}) the replacements
\be
{\cal F}_c \; \rightarrow \; {\cal F}_d, \quad\quad \overline{p}
\;  \leftrightarrow \; -p, \quad\quad k_2 \; \rightarrow \; k_1
\label{eq:e11}
\ee
and changing the overall sign.  Note, however, that ${\cal F}_d =
{\cal F}_c = {\cal F}$, see (\ref{eq:a49}), and that $J^\mu
(k_g)$ (\ref{eq:i3}) remains unchanged after these operations.

So, if we were to omit the $\varepsilon_{\mu \nu \rho \sigma}$
terms, we see that the total contribution of the \lq\lq hard"
quark-quark and Compton subprocesses would
have a simple factorized form given by the product of three
factors:  the Born matrix element of the basic process
$\gamma\gamma \rightarrow q\overline{q}$, the one-loop correction
${\cal F}_i$ coming from the diagrams 4(b--d), and the factor $g_S \langle t^a
\rangle$ $e_\mu (k_g) J^\mu$ for the
accompanying gluon bremsstrahlung.  But it now seems at first
sight that the presence of the $\varepsilon_{\mu \nu \rho
\sigma}$ terms will destroy even the factorization of the Born
amplitude of the basic process.  However, we will find that this
is not the case.

A second apparent problem is that the regions of
quasi-collinearity of the emitted gluon with the momenta of the
initial photons give singularities in the separate contributions
(\ref{eq:e8}), (\ref{eq:e10}) and (\ref{eq:e11}).  However, these
singularities cancel
when we take the sum of the contributions, as one can naively
expect from the physical point of view.  The cancellation can be
demonstrated by using the general generic expression
\be
f (p_1, p_2, e, k_g) \; = \; \frac{\varepsilon_{\mu \nu
\rho \sigma} \: p_1^\mu \: p_2^\nu \: e^\rho \:
k_g^\sigma}{(k_g . p_1)(k_g . p_2)} \: \ln^2 \left ( \frac{(k_g .
p_1)(k_g . p_2)}{m_q^2 (p_1 . p_2)} \right ),
\label{eq:e12}
\ee
where the momenta $p_1, p_2$ are $k_2, k_1$ in (\ref{eq:e8}),
$\overline{p}, k_2$ in (\ref{eq:e10}) and $k_1, p$ in
(\ref{eq:e11}).  Now in the region where the gluon emission is
quasi-collinear with $p_1$, say, that is where 
{\rm angle} ({\boldmath $k_g,p_1$}) $\ll$ angle ({\boldmath $p_2,
p_1$}) it is easy to show that
\be
f (p_1, p_2, e, k_g) \; \simeq \; - \:
\frac{(\mbox{\boldmath $e$} \: \times \: \hat{\mbox{\boldmath $k$}}_g) .
\mbox{\boldmath $p$}_1}{(k_g . p_1)} \: \ln^2
\left ( \frac{(k_g . p_1) \omega_g}{m_q^2 \: E_1} \right ),
\label{eq:e13}
\ee
where $\hat{\mbox{\boldmath $k$}}_g \equiv \mbox{\boldmath $k$}_g/\omega_g$.  Note
that $f$ in (\ref{eq:e13}) is independent
of $p_2$.  Moreover note the equality of the coefficient
functions of ${\cal F}_i$ in the uncrossed terms in
(\ref{eq:a41}) for $M_b$, in (\ref{eq:a54}) for $M_c$ and the
analogous equation for $M_d$.  The proof follows as a by-product
of the demonstration of the equality of the sum of the crossed
and uncrossed terms used in derivation of factorization formula
(\ref{eq:a36}).  Using the above properties we can combine
together the corrections (\ref{eq:e7}), (\ref{eq:e8}),
(\ref{eq:e10}) and (\ref{eq:e11}) to obtain
\bea
\label{eq:e15}
M_{\rm 1-loop}^{(1)} & = & M_{\rm Born} \: g_S \langle t^a
\rangle \: \left [
e_\mu (k_g) \: J^\mu \left \{ {\cal F}_a \: + \: 3{\cal F} \:
- \: \frac{\alpha_S}{8 \pi} \: C_A \: \ln^2 \left (
\frac{(k_g . p)(k_g . \overline{p})}{m_g^2 (p .
\overline{p})} \right ) \right \} \right . \nonumber \\
& & \nonumber \\
& + & i \lambda \: \left . \frac{\varepsilon_{\mu \nu \rho
\sigma} \: \overline{p}^\mu \: p^\nu \: e^\rho (k_g) \:
k_g^\sigma}{(k_g . \overline{p})(k_g . p)} \: \left ( \frac{-
\alpha_S}{4 \pi} \right ) \: \left ( C_F \: - \: \frac{C_A}{2}
\right ) \: \ln^2 \left ( \frac{(k_g . p)(k_g .
\overline{p})}{m_q^2 \: (p . \overline{p})} \right ) \right ],
\eea
where $m_g$ and $\lambda$ denote respectively the gluon
mass and $\gamma\gamma \rightarrow q\overline{q}$ particle
helicities.  The absence of the singularities when the gluon
becomes collinear with either of the initial photons is now
clearly 
manifest.

The presence of the $\varepsilon_{\mu \nu \rho \sigma}$ term in
the one-loop correction to the gluon emission amplitude
(\ref{eq:e15}) demonstrates the non-triviality of physical
phenomena of helicity-violating processes.  Fortunately this term
does not contribute in the approximation (\ref{eq:e1}) that we
are studying.  It is pure imaginary with respect to the Born
amplitude (\ref{eq:i2}) for $\gamma\gamma \rightarrow
q\overline{q}g$ and so it makes no contribution to $d \sigma_{\rm
1-loop} (\gamma\gamma \rightarrow q\overline{q}g)$ of
(\ref{eq:e1}).  This component of the cross section is therefore
given by
\bea
\label{eq:e16}
& & d\sigma_{\rm 1-loop} (\gamma\gamma \rightarrow
q\overline{q}g) \; = \; d\sigma_{\rm Born} (J_z = 0) \:
\frac{\alpha_S C_F}{4 \pi^2}
\nonumber \\
& & \nonumber \\
& & \times \; \int_{\Omega_R} \: \frac{d^3 k}{\omega} \: \frac{2
(p . \overline{p})}{(k . p)(k . \overline{p})} \: \left [ 2 \:
({\cal F}_a \: + \: 3{\cal F}) \: - \: \frac{\alpha_S}{4 \pi} \:
C_A \: \ln^2 \left ( \frac{(k . p)(k . \overline{p})}{m_g^2 (p .
\overline{p})} \right ) \right ].
\eea

Note that the last term of (\ref{eq:e16}) precisely cancels the
contribution (\ref{eq:e5}) from the emission of two gluons with
strongly correlated momenta that we discussed in section 5.1. 
Therefore summing up the contributions (\ref{eq:e3}),
(\ref{eq:e5}) and (\ref{eq:e16}) of the inelastic processes to
the cross section (\ref{eq:e1}) we obtain the following
remarkably simple result
\be
d\sigma_{\rm Born} (\gamma\gamma \rightarrow q\overline{q}gg) \:
+ \: d \sigma_{\rm 1-loop} (\gamma\gamma \rightarrow
q\overline{q}g) \; = \; d\sigma_{\rm Born} (J_z = 0) \: \left [
{\textstyle \frac{1}{2}} \: \delta_R^2 \: + \: 2 \delta_R \:
({\cal F}_a \: + \: 3{\cal F}) \right ],
\label{eq:e17}
\ee
where $\delta_R$ is given by (\ref{eq:i4}) (or, more precisely,
(\ref{eq:e54}), (\ref{eq:g54}) or (\ref{eq:h56})), ${\cal F}_a$
by (\ref{eq:a38}) and ${\cal F} = {\cal F}_b = {\cal F}_c = {\cal
F}_d$ by (\ref{eq:a42}). \\

\noindent {\bf 5.3~~The non-radiative two-loop
contribution}

We now come to the last term in (\ref{eq:e1}), namely the two-loop
contribution, $d\sigma_{2-loop} (\gamma \gamma \rightarrow q 
\overline{q})$, to the cross section of the basic process.  It
consists of two pieces.  The first is the square of the one-loop
corrections, $(M_a + M_b + M_c + M_d)$, to the basic matrix
element
\be
d \sigma_{\rm 2-loop}^{(1)} (\gamma \gamma \rightarrow q\overline{q}) \; =
\; d\sigma_{\rm Born} (J_z = 0) \left ( \sum_i {\cal F}_i \right
)^2,
\label{eq:f1}
\ee
see (\ref{eq:a36}).  The ${\cal F}_i$ are given  by (\ref{eq:a38}), (\ref{eq:a42}),
(\ref{eq:a48}) and (\ref{eq:a49}).

The second piece comes from the interference of the two-loop
correction to the matrix element with its Born value (\ref{eq:a34}).  The
calculation can be performed in a similar way to that used in the
previous section.  Again we separate the \lq \lq hard" and \lq
\lq soft" stages of the process.  The soft stage is the source
of the DL contributions whereas, by dimensional arguments, we
find that the hard stage is a two-to-two process.  Therefore we
have the same possibilities as for the one-loop correction (see
Fig.~4).  The DL contributions can come from either a soft gluon
or a soft quark, but the quark can only occur once since it leads
to a $m_q/M_H$ suppression of the amplitude (which is the usual 
suppression of the helicity violating amplitudes).  Therefore,
two-loop DL corrections can be obtained from the diagrams of
Fig.~4 by the insertion of a soft gluon line.  We call these soft
insertions.  Of course we do not have to consider a soft gluon
emitted from the \lq \lq hard blob".

If the hard process is $\gamma \gamma \rightarrow q \overline{q}$
then its dominant amplitude
$M_{\lambda,\lambda}^{\lambda,\lambda}$ is already
suppressed by $m_q/M_H$ (see (\ref{eq:a7})), and so
the DL contributions come only from soft gluons.  Due to the
factorization of the hard part of the matrix element this case is
exactly analogous to the quark form factor.  The diagrams are
obtained by soft insertions in Fig.~4(a).  In the Feynman gauge
(which is used for virtual gluons) the DL contributions
only occur when the soft gluon connects lines with strongly
different momenta, that is momenta $p_i$ and $p_j$ which satisfy
\be
|p_i \cdot p_j| \; \gg \; |p_i^2|, |p_j^2|.
\label{eq:f2}
\ee
The contributing diagrams, shown in Fig.8, give in total
\be
M_{\rm 2-loop}^{\rm Fig.8} \; = \; M_{\rm Born} \: \frac{{\cal
F}_a^2}{2}.
\label{eq:f3}
\ee
If the hard process is $q \overline{q} \rightarrow q \overline{q}$ then we 
must make soft insertions in Fig.~4(b).  Since the DL contributions only occur
when the soft gluon connects line with strongly different momenta we need only 
consider the diagrams shown in Fig.~9.  We discuss the details in Appendix B.  
The final result is that diagrams 9(g) and 9(h) do not contribute, while 9(c) -
9(f) cancel each other.  Therefore the net contribution comes only from
diagrams 9(a,b) and is equal to
\be
M_{\rm 2-loop}^{\rm Fig.9} \; = \; M_{\rm Born} \: {\cal F}_b \left
( {\cal F}_a \: + \: \frac{{\cal F}_b}{6} \right ).
\label{eq:f4}
\ee
Note that the second term, proportional to ${\cal F}_b^2$, coincides with the 
one-loop QCD correction to the light quark contribution to the $H \rightarrow
\gamma \gamma$ decay amplitude presented in \cite{SDGZ}, see also \cite{KYA}.

For the case when the hard process is $\gamma q\rightarrow qg$ the relevant
soft gluon insertions are shown in the diagrams of Fig.~10.  Their individual
contributions are given in Appendix B.  The total result is
\be
M_{\rm 2-loop}^{\rm Fig.10} \; = \; M_{\rm Born} \: {\cal F}_c
\left ({\cal F}_a \: + \: \frac{C_A}{2 C_F} \: \frac{{\cal
F}_c}{6} \right ).
\label{eq:f5}
\ee
The contributions from the $\gamma \overline{q} \rightarrow \overline{q} g$ hard
process give the same result, when we take into account (\ref{eq:a49}).

In summary, the complete two-loop correction to the matrix element of the basic
process is given by the sum of (\ref{eq:f3}), (\ref{eq:f4}) and twice (\ref{eq:f5}).
Using (\ref{eq:a42}) and (\ref{eq:a48}) we have
\be
M_{\rm 2-loop} \; = \; M_{\rm Born} \left [ {\cal F}_a \: \left (
\frac{{\cal F}_a}{2} \: + \: 3 {\cal F} \right ) \: + \: \left (1
\: + \: \frac{C_A}{C_F} \right ) \: \frac{{\cal F}^2}{6} \right
].
\label{eq:f6}
\ee
Recall that the two-loop contribution $d \sigma_{2-{\rm loop}} (\gamma \gamma
\rightarrow q\overline{q})$ is the sum of two pieces, namely the sum
of (\ref{eq:f1}) and 
\be
d \sigma_{\rm 2-loop}^{(2)} \: (\gamma \gamma \rightarrow q\overline{q}) \;
= \; 2 {\rm Re} (M_{\rm Born}^* M_{\rm 2-loop}).
\label{eq:f7}
\ee
Thus we have in total
\be
d \sigma_{\rm 2-loop} \: (\gamma \gamma \rightarrow q\overline{q}) \; = \;
d\sigma_{\rm Born} (J_z = 0) \: \left [ 2 {\cal F}_a ({\cal F}_a
+ 6 {\cal F}) \: + \: {\cal F}^2 \left (9 \: + \: \frac{1}{3} \:
\left ( 1 \: + \: \frac{C_A}{C_F} \right ) \right ) \right ],
\label{eq:f8}
\ee
where ${\cal F}_a$ is given by (\ref{eq:a38}) and ${\cal F}$ by (\ref{eq:a42}). \\

\noindent {\bf 5.4~~The non-Sudakov form factor $F_q$}

The above results allow us to obtain the two-loop contribution to the
non-Sudakov form factor $F_q$.  We substitute (\ref{eq:e17}) and (\ref{eq:f8}) into
(\ref{eq:e1}), and use the representation
\be
d \sigma_{\rm 2-loop} \; = \; d \sigma_{\rm Born} \: (J_z = 0) \:
(F_g F_q)_2,
\label{eq:f9}
\ee
where the subscript 2 indicates that we should take the $\alpha_s^2$ terms in the
expansion of the product of the two form factors ($F_g$ is given by (\ref{eq:i5}), 
and the one-loop approximation to $F_q$ is given in (\ref{eq:i6})).  In this way
we obtain
\bea
\label{eq:f10}
F_q & = & 1 \: + \: 6 {\cal F} \: + \: {\cal F}^2 \left (9 \: +
\: \frac{1}{3} \: \left (1 \: + \: \frac{C_A}{C_F} \right )
\right ) \nonumber \\
& & \nonumber \\
& = & (1 + 3 {\cal F})^2 \: + \: \frac{{\cal F}^2}{3} \: \left (1
\: + \: \frac{C_A}{C_F} \right ).  
\eea 

\noindent {\large \bf 6.~~Summary and discussion}

We have studied Higgs production in polarised $\gamma \gamma$ collisions.
In particular we have investigated the feasibility of the proposal that the 
Higgs may be isolated
in the $\gamma \gamma \: (J_z = 0) \rightarrow b \overline{b}$ channel, due to 
the remarkable $m_q^2 / s$ suppression of the 
background process $\gamma \gamma \rightarrow q \overline{q}$. However, the
especially large
radiative corrections to the background process cause the situation to be much 
more complicated than it appears at first sight.  Indeed it is essential to 
perform a detailed analysis of the various (real and
virtual) background processes since these can greatly exceed the leading order
(Born) $\gamma \gamma \rightarrow q \overline{q}$ result.

The general structure of the background arising from the central production of
quasi-two-jet-like events with at least one tagged energetic $b$ jet was written
in the form
\be
\sigma (\gamma \gamma \rightarrow 2 \, {\rm jets}) \; = \; \sigma (\gamma \gamma
\rightarrow q \overline{q}) F_g F_q \; + \; \sigma (\gamma \gamma \rightarrow
q \overline{q} g \rightarrow 2 \, {\rm jets})_{{\rm collinear + Compton}}
\label{eq:s1}
\ee
see (\ref{eq:a19}).  Here it is to be understood that the
incoming $\gamma \gamma$ system is in the $J_z = 0$ state.  A major problem is 
that radiative $q \overline{q} g$ production in the collinear and Compton 
configurations do not have the $m_q^2 / s$ suppression, and so could exceed the Born
estimate of the background.  In the Compton configuration we are concerned with 
the production of a comparatively soft $q$ or $\overline{q}$ which goes 
undetected along the beam direction.  The contribution was calculated in Section
3 and was found to be quite sizeable, see eqs. (\ref{eq:b32}) - (\ref{eq:d32}).
Therefore it should be avoided, if at all possible, by tagging both the energetic
$b$ and $\overline{b}$ jets.

Collinear gluon bremsstrahlung off one of quarks was investigated in \cite{BKSO}. 
It can be suppressed by using traditional cuts to discriminate between the two
and three jet topologies, but then due to the Sudakov form factor $F_g$ we also
deplete the signal.  The Sudakov form factor $F_g$ was given in (\ref{eq:f54}),
(\ref{eq:h54})
and (\ref{eq:h57}) for three different cut-off prescriptions.

One of our main aims has been the calculation of the non-Sudakov form factor
$F_q$.  It involves novel double logarithmic (DL) terms. Our result, in the
two-loop 
approximation, is shown in (\ref{eq:f10}).  Let us summarize the structure of 
this form factor.

In the one-loop approximation $F_q$ is given by (\ref{eq:i6}).  The crucial
observation is that the coefficient $c_1$ in the expansion (\ref{eq:a21}) of 
$F_q$ in powers of $(\alpha_S / \pi) L_m^2$ is anomalously large and negative,
$c_1 = -8$.  Thus the cross section, calculated to order $\alpha_S$ accuracy,
could formally become negative \cite{JT1}.  At first sight this indicates that it is
necessary to sum the whole series.  Fortunately we find that it is not the case.
The calculation of $F_q$ in the two-loop approximation (\ref{eq:f10}) is quite
sufficient, and shows that the higher order coefficients are not so anomalously
large.  Moreover the large size of $c_1$ has a simple physical explanation.
Recall that $F_q$ is specific to the helicity-violating process and arises from
the soft quark contributions.  It is not connected with soft real gluon emission
which,
together with the soft gluon virtual contribution, is absorbed in $F_g$.  Now, the
soft quark corrections to the matrix element $M_{\lambda \lambda}^{\lambda \lambda}$
of the basic $\gamma \gamma \rightarrow q \overline{q}$ process come from three
different kinematical configurations (or, equivalently are connected with
three hard subprocesses, see Figs 4(b) - 4(d)).  We thus have a factor 3
enhancement of the
amplitude and a factor 6 in the cross section.  At higher orders the essential 
point is that the number of hard subprocesses remains
the same.  Thus there is a loss of a factor 3 in the two-loop correction to the
amplitude as compared with that estimated by the square by the first-order 
correction.  Moreover, the higher-order corrections to $F_q$ arise from 
kinematical regions where there is one soft quark but several soft gluons, which
nevertheless have to be harder than the soft quark since otherwise they are 
absorbed in $F_g$.  This requirement reduces the higher order coefficients.
Equation (\ref{eq:f6}) offers a good example of the above effects. Here we
have only to consider the term containing ${\cal F}^2$
because the terms involving ${\cal F}_a$ are absorbed in ${\cal F}_g$.  
Roughly speaking in this case we have a factor $(1 + C_A / C_F) \simeq 3$
corresponding to the number of kinematic regions, while the contribution of
each region has a coefficient 1/6 due to the restrictions on the region of 
integration over the soft gluon. 

In summary, we have presented a full study of the important radiative effects
accompanying $\gamma \gamma \rightarrow {\rm two}$ heavy-quark-jets (in the $J_z = 0$
channel).  The aim has been to estimate the contributions from the various
radiative background processes to the $H \rightarrow {\rm two} \, b$-jet signal.
A more quantitative study will require improvements on both the
theoretical and the experimental side.  On the theoretical
front we will need a self consistent analysis of the single logarithmic terms
including the effects of the running mass in the background processes and of the
evaluation of the running coupling in both the Sudakov and non-Sudakov form
factors.  One of the most important experimental questions is the efficiency
of $b$-tagging and of the rejection of $c \overline{c}$ events.  There are
various different estimates of the level of $c \overline{c}$ contamination,
see \cite{BBC,BKSO,OTW}.  Another important task is to find the optimal choice of cut-off prescription to define
the two-jet configurations.  Finally we note that for an intermediate mass Higgs,
say $M_H = 100 {\rm GeV}$, we find, using (\ref{eq:a42}) with
$\alpha_S (M_H) = 0.12$, that 
${\cal F} \approx -0.5 \, ({\rm and} -0.9)$ for $b \overline{b}$ (and $c \overline{c}$)
production.  From (\ref{eq:f10}) we see that this gives a factor of about
3 suppression in the first term in (\ref{eq:s1}) or (\ref{eq:a19}), which qualitatively
justifies the expectations of \cite{BKSO}.  At the same time, the potential
$c \overline{c}$ non-radiative contribution is enhanced by a factor of about 4.

\noindent {\large \bf Acknowledgements}

We thank J.\ Campbell, E.W.N.\ Glover and W.J.\ Stirling for discussions.
We also thank the UK Particle Physics and Astronomy Research Council for financial
support.  VSF thanks Grey College of the University of Durham for their warm
hospitality and the support of the Russian Fund for Basic Research, as well as 
INTAS grant 95-0311.

\newpage
\noindent {\large \bf Appendix A}

Here we derive the formulae for the one-loop corrections to the
amplitude for the process $\gamma\gamma \rightarrow
q\overline{q}g$ in the cases of the \lq\lq hard" subprocesses
$q\overline{q} \rightarrow q\overline{q}$ and $\gamma q
\rightarrow qg$.  The corresponding diagrams are shown in
Figs.~6,7.  Note that some of these diagrams have contributions
which are singular when the emitted gluon is collinear with
either of the initial photons.  Therefore they will not
contribute to the cross section (\ref{eq:e1}), because in the
Born approximation we only have collinear singularities when the
gluon is emitted in the direction of either the outgoing quark or
antiquark.  Nevertheless it is useful to evaluate them because
each diagram separately gives a contribution, which could lead to
double logarithmic terms at ${\cal O} (\alpha_S^3)$.  Moreover we
meet novel double logarithmic terms on account of the
helicity-violating nature of the basic process.

We start with diagram 6(a).  The contribution of this diagram and
the diagram 6($\overline{\rm a}$), in which the gluon is emitted
from the $\overline{q}$, can be immediately written down, since
the soft gluon cannot influence the hard process.  We have
\renewcommand{\theequation}{A.1}
\be
M_{\rm 1-loop}^{\rm (1) Fig.6a,\overline{a}} \; = \; M_{\rm Born}
\: {\cal F}_b \: g_S \langle t^a \rangle \: e_\mu (k_g) \: J^\mu
(k_g).
\label{eq:e18}
\ee
Diagram 6(b) can be evaluated using standard
techniques.  We first neglect the momentum $k_g$ of the soft
gluon in the numerator of the matrix element.  Now recall that
the DL contribution comes from the region where the momentum
of the quark which emits the gluon is nearly equal to $k_1$. 
Thus we can reduce the numerator to its value without gluon
emission multiplied by $2g_S (e (k_g) . k_1)$.  In this
way we obtain a factorized form similar to (\ref{eq:a41}) but
with the replacement
\renewcommand{\theequation}{A.2}
\be
{\cal F}_b \; \rightarrow \; \frac{2 g_S (e (k_g) .
k_1)}{- 2 k_1 . k_g} \: \frac{(C_F - {\textstyle \frac{1}{2}}
C_A)}{C_F} \: \langle t^a \rangle \: {\cal F}_{6b},
\label{eq:z1}
\ee
where ${\cal F}_{6b}$ only differs from ${\cal F}_b$ by the
restriction that
\renewcommand{\theequation}{A.3}
\be
| k_1 . k | \; \ll \; k_1 . k_g
\label{eq:z2}
\ee
on the region of the $k$ integration in (\ref{eq:a42}).  This
restriction appears because the contribution from outside the
region does not have DL behaviour due to the additional
propagator in diagram 6(b) as compared to diagram 4(b).  We use
Sudakov variables defined by
\renewcommand{\theequation}{A.4}
\be
k \; = \; \beta k_1 \: + \: \alpha k_2 \: + \: k_\perp
\label{eq:z3}
\ee
to evaluate ${\cal F}_{6b}$.  Then restriction (\ref{eq:z2})
becomes
\renewcommand{\theequation}{A.5}
\be
| \alpha | \: \ll \: | \alpha_g | \; \equiv \; \frac{k_g .
k_1}{k_2 . k_1}
\label{eq:z4}
\ee
and
\renewcommand{\theequation}{A.6}
\bea
\label{eq:z5}
{\cal F}_{6{\rm b}} & = & - \: \frac{\alpha_S}{2 \pi} \: C_F \:
\int_0^1 \int_0^1 \: \frac{d\alpha}{\alpha} \:
\frac{d\beta}{\beta} \: \theta (\alpha \beta s - m_q^2) \: \theta
\left ( \frac{2 k_1 . k_g}{s} \: - \: \alpha \right ) \nonumber
\\
& & \nonumber \\
& = & - \: \frac{\alpha_S}{4 \pi} \: C_F \: \ln^2 \left (
\frac{2 k_1 . k_g}{m_q^2} \right ),
\eea
where it is assumed that $s \gg 2 k_1 . k_g \gg
m_q^2$.

In a similar way it is easy to see that diagram 6(c) has the form
(\ref{eq:a41}) with the replacement
\renewcommand{\theequation}{A.7}
\be
{\cal F}_b \; \rightarrow \; \frac{- 2 g_S (e (k_g) .
k_2)}{- 2 k_g . k_2} \: \frac{(C_F - {\textstyle \frac{1}{2}}
C_A)}{C_F} \: \langle t^a \rangle \: {\cal F}_{6c}
\label{eq:z6}
\ee
where
\renewcommand{\theequation}{A.8}
\be
{\cal F}_{6{\rm c}} \; = \; - \: \frac{\alpha_S}{4 \pi} \: C_F \:
\ln^2 \left ( \frac{2 k_2 . k_g}{m_q^2} \right ),
\label{eq:z7}
\ee
with the limitation that $s \gg 2 k_2 . k_g \gg m_q^2$.

The evaluation of the contribution of Fig.~6(d) is more
complicated due to its spin
structure.  The result is quite novel.  Just as in the original
\lq elastic' amplitude with \lq\lq hard" subprocess
$q\overline{q} \rightarrow q\overline{q}$ of diagram 4(b), the
numerators of the $q$ and
$\overline{q}$ propagators entering the \lq\lq hard blob" are
well approximated by $\fmslash{k}_1$ and $\fmslash{k}_2$.  We use
for them the representation (\ref{eq:a40}).  However,
for the quark line between the photon vertices we have to replace
$m_q$ by
\renewcommand{\theequation}{A.9}
\be
m_q \left [ \fmslash{k} \fmslash{e} (k_g) \: + \:
\fmslash{e} (k_g) (\fmslash{k} + \fmslash{k}_g) \right
] \; = \; m_q \left [ 2 k . e (k_g) \: + \:
\fmslash{e} (k_g) \fmslash{k}_g \right ]
\label{eq:z8}
\ee
where $k$ is the quark momentum shown in Fig.~6(d).  The spin
structure of the first term is the same as the \lq elastic' case,
but the second term has to be treated separately.  Using
(\ref{eq:a39}) we find we have to calculate the spin matrix
element
\renewcommand{\theequation}{A.10}
\be
{\cal S} \; \equiv \; \overline{u}^\lambda (k_1) \:
\fmslash{e}^\lambda (k_1) \: \fmslash{e} (k_g) \:
\fmslash{k}_g \: \fmslash{e}^\lambda (k_2) \: v^\lambda (k_2).
\label{eq:z9}
\ee
This matrix element is gauge invariant with respect to the gluon,
as well as to the photons.  We can reduce it to the \lq elastic'
factorized form (\ref{eq:a41}) by decomposing the four vectors
$e (k_g)$ and $k_g$ in terms of $k_1, k_2, e_1$ and
$e_2$ (where $e_i \equiv e^\lambda(k_i)$) which satisfy
\renewcommand{\theequation}{A.11}
\bea
k_1^2 & = & k_2^2 \; = \; e_1^2 \; = \; e_2^2 \; = \; 0,
\nonumber \\
k_i . e_j & = & 0 \quad \hbox{for} \quad i, j \; = \; 1,2,
\nonumber
\eea
provided that we choose a gauge in which $k_1 . e_2 = k_2 . e_1 =
0$.  After simple Dirac algebra we find
\renewcommand{\theequation}{A.12}
\be
{\cal S} \; = \; \left [ \frac{2 (e . k_1)(k_g .
k_2)}{(k_1 . k_2)} \: + \: \frac{2 (e . e_1)(k.
e_2)}{(e_1 . e_2)} \right ] \: \overline{u}^\lambda (k_1) \:
\fmslash{e}^\lambda (k_1) \: \fmslash{e}^\lambda (k_2) \:
v^\lambda (k_2),
\label{eq:z10}
\ee
which has the original spin structure of (\ref{eq:a41}).  We now
use representation (\ref{eq:a35}) for the polarisation
vectors $e_i$ to rewrite the second term in the square brackets
in the form
\renewcommand{\theequation}{A.13}
\bea
\label{eq:z11}
\frac{2 (e . e_1)(k_g . e_2)}{(e_1 . e_2)} & = & -
(\mbox{\boldmath $e$} . \mbox{\boldmath $k$}_g) \: +
\: i \lambda (\mbox{\boldmath $e$} \times
\mbox{\boldmath $k$}_g) . \hat{\mbox{\boldmath $k$}}_1 \nonumber \\
& & \nonumber \\
& = & \frac{- e . (k_1 (k_g . k_2) \: + \: k_2
(k_g . k_1))}{k_1 . k_2} \: + \: i\lambda
\frac{\varepsilon_{\mu \nu \rho \sigma} \: k_2^\mu \: k_1^\nu \:
e^\rho \: k_g^\sigma}{k_1 . k_2}
\eea
where $\hat{\mbox{\boldmath $k$}}_i$ is a unit 3-vector and where the $\gamma,
\gamma, q, \overline{q}$ helicities are all given by $\lambda$,
and $\varepsilon_{0123} = 1$ etc.  Thus the contribution of
diagram 6(d) is obtained from (\ref{eq:a41}) with the replacement
\renewcommand{\theequation}{A.14}
\be
{\cal F}_b \; \rightarrow \; 16g_S (C_F \: - \: {\textstyle
\frac{1}{2}} C_A) \: \langle t^a \rangle \: \left ( \frac{-
\alpha_S}{4 \pi} \right ) \: I,
\label{eq:z12}
\ee
where the integral
\renewcommand{\theequation}{A.15}
\bea
\label{eq:z13}
&I&  =  \nonumber \\
\nonumber \\
&\displaystyle\int&\frac{d^4 k}{i (2 \pi)^2} \:
\frac{e . (2k (k_1 . k_2) \: + \: k_1 (k_g .
k_2) \: - \: k_2 (k_g . k_1)) \: + \: i\lambda \:
\varepsilon_{\mu \nu \rho \sigma} \: k_2^\mu \: k_1^\nu \:
e^\rho \: k_g^\sigma}{[(k + k_1)^2 - m_q^2 +
i\varepsilon][(k + k_g - k_2)^2 - m_q^2 + i\varepsilon][(k +
k_g)^2 - m_q^2 + i \varepsilon][k^2 - m_q^2 + i\varepsilon]} \nonumber
\\
\eea
This integral may be evaluated using the Feynman parameter
technique.  We define $x_i$ to be the Feynman parameter for the
$i$th denominator in the integrand.  We perform the exact
integration over $x_4$ and $x_3$, and obtain
\renewcommand{\theequation}{A.16}
\bea
\label{eq:z13a}
I & = & \frac{1}{4} \int_0^1 \int_0^1 \: dx_1 dx_2 (1 - x_1 -
x_2) \:
\theta (1 - x_1 - x_2) \nonumber \\
& & \\
& \times & \frac{(e . k_1)(k_g . k_2 - 2x_1
k_1 . k_2) \: - \: (e . k_2)(k_g . k_1 - 2x_2 k_1 .
k_2) \: + \: i \lambda \varepsilon_{\mu \nu \rho \sigma} \:
k_2^\mu \: k_1^\nu \: e^\rho \: k_g^\sigma}{D_1 D_2} \nonumber
\eea
with
$$
D_i \; \equiv \; - 2x_1 x_2 (k_1 . k_2 - k_j . k_g) \: + \: 2x_i
(1 - x_i) \: k_i . k_g \: + \: m_q^2 \: - \: i \varepsilon
$$
where $k_j = k_1$ if $i = 2$ and $k_j = k_2$ if $i = 1$.  We now
extract the double logarithmic behaviour from this integral
form and find
\renewcommand{\theequation}{A.17}
\be
16I \; = \; \frac{e . k_1}{k_g . k_1} \: \ln^2 \left
( \frac{2 k_g . k_1}{m_q^2} \right ) \: - \: \frac{e . k_2}{k_g
. k_2} \: \ln^2 \left ( \frac{2 k_g . k_2}{m_q^2} \right
) \: + \: i \lambda \frac{\varepsilon_{\mu \nu \rho \sigma}
k_2^\mu \: k_1^\nu \: e^\rho \: k_g^\sigma}{(k_g .
k_1)(k_g . k_2)} \: \ln^2 \left ( \frac{(k_g . k_1)(k_g .
k_2)}{m_q^2 \: (k_1 . k_2)} \right ),
\label{eq:z14}
\ee
where it has been assumed that the arguments of all the
logarithms are large\footnote{More precisely each logarithm is
only taken into account when its argument is large.  This
assumption is made throughout this Appendix.}.

If we now combine together all the contributions coming from the
diagrams of Fig.~6 we obtain (\ref{eq:e8}). Note that all terms
proportional to ($e . k_i$) cancel each other.

Now we consider the contribution of the diagrams of Fig.~7, which
arise by adding a soft gluon line to diagram 4(c).  First, the
contribution of diagram 7(a), together with the crossed diagram
with $k_1 \leftrightarrow k_2$, has the factorised form
\renewcommand{\theequation}{A.18}
\be
M_{\rm 1-loop}^{(1) \: {\rm Fig.7a}} \; = \; M_{\rm Born} \: g_S
\langle t^a \rangle \: \frac{e . p}{k_g . p} \: {\cal F}_c.
\label{eq:z15}
\ee
In a similar way the contribution of diagram 7(b) has the form
(\ref{eq:a54}) with the replacement
\renewcommand{\theequation}{A.19}
\be
{\cal F}_c \; \rightarrow \; - g_S \langle t^a \rangle \:
\frac{e . \overline{p}}{k_g . \overline{p}} \: {\cal
F}_{7b},
\label{eq:z16}
\ee
but now ${\cal F}_{7b}$ is obtained from ${\cal F}_c$ with the
limitation
\renewcommand{\theequation}{A.20}
\be
| k . \overline{p} | \; \gg \; k_g . \overline{p}
\label{eq:z17}
\ee
on the $k$ integration in (\ref{eq:a46}).  Thus we have
\renewcommand{\theequation}{A.21}
\be
{\cal F}_{7{\rm b}} \; = \; - \frac{\alpha_S}{4 \pi} \: C_F \left [
\ln^2 \left ( \frac{2 k_2 . \overline{p}}{m_q^2} \right ) \: - \:
\ln^2 \left ( \frac{2 k_g . \overline{p}}{m_q^2} \right ) \right
],
\label{eq:z18}
\ee
where, as always, we assume that the arguments of the logarithms
are large.

The contributions of diagrams 7(c,d) can be calculated in a
similar way to that used for diagrams 6(b,c).  Thus they have the
form of (\ref{eq:a54}) with the following replacements
respectively
\renewcommand{\theequation}{A.22}
\bea
\label{eq:z19}
{\cal F}_c & \rightarrow & -g_S \langle t^a \rangle \: (C_F -
{\textstyle \frac{1}{2}} C_A) \: \frac{e . k_2}{k_g .
k_2} \: \left ( \frac{- \alpha_S}{4 \pi} \right ) \: \ln^2 \left
( \frac{2 k_g . k_2}{m_q^2} \right ), 
\eea
\renewcommand{\theequation}{A.23}
\bea
\label{eq:z20}
{\cal F}_c & \rightarrow & -g_S \langle t^a \rangle \:
{\textstyle
\frac{1}{2}} C_A \: \frac{e . \overline{p}}{k_g .
\overline{p}} \: \left ( \frac{- \alpha_S}{4 \pi} \right ) \:
\ln^2 \left ( \frac{2 k_g . \overline{p}}{m_q^2} \right ).
\eea
Finally diagram 7(e) may be evaluated in a similar way to diagram
6(d).  Again its contribution is the form of (\ref{eq:a54}), but
now with the replacement
\renewcommand{\theequation}{A.24}
\bea
\label{eq:z21}
{\cal F}_c & \rightarrow & g_S \langle t^a \rangle \: (C_F -
{\textstyle \frac{1}{2}} C_A) \: \left ( \frac{- \alpha_S}{4 \pi}
\right ) \: \left [ \frac{e . k_2}{k_g . k_2} \:
\ln^2 \left ( \frac{2 k_g . k_2}{m_q^2} \right ) \right .
\nonumber \\
& & \nonumber \\
& - & \left . \frac{e . \overline{p}}{k_g .
\overline{p}} \: \ln^2 \left ( \frac{2 k_g . \overline{p}}{m_q^2}
\right ) \: + \: i \lambda \frac{\varepsilon_{\mu \nu \rho
\sigma} \: \overline{p}^\mu \: k_2^\nu \: e^\rho \:
k_g^\sigma}{(k_g . k_2)(k_g . \overline{p})} \: \ln^2 \left (
\frac{(k_g . \overline{p})(k_g . k_2)}{m_q^2 \: (k_2 .
\overline{p})} \right ) \right ].
\eea
Combining together all the above contributions of Fig.~7 we
finally obtain (\ref{eq:e10}).  We see the cancellations of the
spurious $k_g . k_1$ and $k_g . k_2$ singularities.  These
singularities in the $\varepsilon_{\mu \nu \rho \sigma}$ term
cancel in the total $M_{\rm 1-loop}^{(1)}$ contribution given in
(\ref{eq:e15}).

\newpage
\noindent {\large \bf Appendix B}

Here we derive the DL corrections, (\ref{eq:f4}) and (\ref{eq:f5}), connected 
with the hard processes $q \overline{q} \rightarrow q\overline{q}$ and
$\gamma q \rightarrow gq$ respectively.  The diagrams for the first
case are shown in Fig.~9.  The contribution of diagram 9(a) can be
written down immediately.  The exchange of the soft quark with momentum $k_q$
occurs before, whereas the exchange of soft gluon with momentum $k_g$ occurs
after, the hard $q \overline{q} \rightarrow q \overline{q}$ scattering.  Thus, 
since the exchanges do not influence each other, we have
\renewcommand{\theequation}{B.1}
\be
M^{{\rm Fig.9a}} \; = \; M_{\rm Born} {\cal F}_a {\cal F}_b.
\label{eq:x1}
\ee
Recall that this amplitude (and those below) represents the sum of the 
contribution of diagram 9(a) and the diagram with the photon momenta
interchanged.

For the calculation of the contribution of diagram 9(b) it is convenient to 
use the Sudakov decomposition of the momenta of the soft particles
\renewcommand{\theequation}{B.2}
\be
k_i \; = \; \beta_i k_1 \: + \: \alpha_i k_2 \: + \: k_{iT},
\label{eq:x2}
\ee
where $i = {\rm (soft)} \; q \; {\rm or} \; g$.  Recall that the DL contributions come
from the regions
\renewcommand{\theequation}{B.3}
\bea
\label{eq:x3}
1 \; \gg \; |\alpha_i|, |\beta_i| \; \gg \; 
|k_{iT}^2 / s| \; \gg \; |m_i^2 / s| 
\eea
and can be calculated performing the integration over the corresponding 
transverse momenta of the soft particles by taking half of the residues in the
corresponding propagators 
\renewcommand{\theequation}{B.4}
\bea
\label{eq:x4}
\frac{d^4 k_i}{k_i^2 \: - \: m_i^2 \: + \: i \varepsilon} & = & 
\left( \frac{s}{2} \right) \: \frac{d \alpha_i \: d \beta_i \: d^2 k_{iT}}
{s \alpha_i \beta_i \: - \: \mbox{\boldmath$k$}_{iT}^2 \: - \: m_i^2 \: + \: i \varepsilon} 
\nonumber \\
& & \\
& & \rightarrow \: -i \pi^2 \left( \frac{s}{2} \right) d \alpha_i d \beta_i \:
\Theta \: (s \alpha_i \beta_i \: - \: m_i^2). \nonumber
\eea
Thus the DL contribution has the form
\renewcommand{\theequation}{B.5}
\be
M_i \; = \; M_{\rm Born} \: {\cal F}_i,
\label{eq:x5}
\ee
where ${\cal F}_i$ are given by integrals over $\alpha_i$ and $\beta_i$
\renewcommand{\theequation}{B.6}
\bea
\label{eq:x6}
{\cal F}_i & = & \left ( \frac{\alpha_S}{2\pi} \right )^2 \:
\int_0^1 \: \int_0^1 \: \frac{d \alpha_q}{\alpha_q} \: \frac{d
\beta_q}{\beta_q} \: \Theta \left ( \alpha_q \beta_q \: - \:
\frac{m_q^2}{s} \right ) \nonumber \\
& & \nonumber \\
& & \times \; \int_0^1 \: \int_0^1 \: \frac{d \alpha_g}{\alpha_g}
\: \frac{d \beta_g}{\beta_g} \: \Theta \left ( \alpha_g \beta_g
\: - \: \frac{m_g^2}{s} \right ) \: C_i,
\eea
The factor $C_i$ includes both the appropriate colour factors and the
restrictions on $\alpha_i$ and $\beta_i$ necessary to ensure that the
matrix element has logarithmic behaviour in each of the variables
$\alpha_i$ and $\beta_i$.  For
the contribution of diagram 9(b) we find
\renewcommand{\theequation}{B.7}
\be
C_{9b} \; = \; C_F^2 \: \Theta (\alpha_g \: - \: \alpha_q)\Theta
(\beta_g \: - \: \beta_q),
\label{eq:x7}
\ee
and therefore we obtain
\renewcommand{\theequation}{B.8}
\be
{\cal F}_{9{\rm b}} \; = \; \frac{1}{6} \: {\cal F}^2,
\label{eq:x8}
\ee
where from (\ref{eq:a42}) we have
\renewcommand{\theequation}{B.9}
\be
{\cal F} \; = \; - \frac{\alpha_S}{\pi} \: C_F \: \ln^2 \left (
\frac{M_H}{m_q} \right ).
\label{eq:x9}
\ee

The DL contributions of the other diagrams can be calculated in a similar way,
provided that we use an appropriate choice of the base vectors of the Sudakov 
decomposition.  For
the soft quark the choice is the same for all the diagrams of Fig.~9.  However,
for the soft gluon it is convenient to use the light cone momenta $k_1$ and
$(p - m_q^2 \: k_1 / 2p \cdot k_1)$ for diagram 9(c), $k_2$ and 
$(\overline{p} - m_q^2 \: k_2 / 2 \overline{p} \cdot k_2)$ for diagram 9(d), 
and so on.  Since we consider large angle $q \overline{q}$ production, we have
\renewcommand{\theequation}{B.10}
\be
2 p \cdot k_1 \; = \; 2 \overline{p} \cdot k_2 \: \sim \: 2 p \cdot k_2 \; = 
\; 2 \overline{p} \cdot k_1 \: \sim \: 2 p \cdot \overline{p} \: \sim \: 
2 k_1 \cdot k_2 \; = \; s
\label{eq:x10}
\ee
and therefore we can neglect the difference between these variables in the arguments
of the logarithms.  Thus for diagrams 9(c) - 9(f) the factors $C_i$ of
(\ref{eq:x6}) can be shown to be
\renewcommand{\theequation}{B.11}
\bea
\label{eq:x11}
C_{9{\rm c}} & = & - C_{9{\rm e}} \; = \; C_F \left (C_F \: - \:
\frac{C_A}{2} \right ) \: \Theta (\alpha_g - \alpha_q) \: \Theta
\left (\beta_g \: - \: \frac{m_q^2}{s} \: \alpha_g \right );
\nonumber \\
& & \\
C_{9{\rm d}} & = & - C_{9{\rm f}} \; = \; C_F \left ( C_F \: - \:
\frac{C_A}{2} \right ) \: \Theta (\beta_g - \beta_q) \: \Theta
\left ( \alpha_g \: - \: \frac{m_q^2}{s} \: \beta_g \right );
\nonumber
\eea
so the DL contributions of these diagrams cancel each other.

It is also evident that the DL contributions of diagrams 9(g) and 9(h) will 
cancel each other, just as, for example, the contributions of 9(c) and 9(e)
cancel each other.  In fact a simple observation shows that 9(g) and 9(h) are
individually zero in the DL approximation.  Indeed, if we perform an integration
over the soft quark momentum for a fixed value of the momentum of the soft 
gluon then we see that the result
is antisymmetric with respect to the replacement $k_1 \leftrightarrow k_2$, 
just as in the real emission case, see (\ref{eq:z5}).  Since the large variables
(\ref{eq:x10}) in the DL factors do not differ we conclude the contributions
of diagrams 9(g) and 9(h) are separately zero.  Thus we are left with the sum of
diagrams 9(a) and 9(b), that is of (\ref{eq:x1}) and (\ref{eq:x8}), which gives
the result stated in (\ref{eq:f4}).

We now turn to the diagrams shown in Fig.~10.  Their DL contributions can be
calculated in a similar way.  We give below the results for the $C_i$ factors
occurring in representation (\ref{eq:x5}), (\ref{eq:x6}) for the individual 
diagrams
\renewcommand{\theequation}{B.12}
\bea
\label{eq:x12}
C_{10{\rm a}} & = & C_F^2 \: \Theta \left (\alpha_g \: - \:
\frac{m_q^2}{s} \: \beta_g \right ) \: \Theta \left (\beta_g \: -
\: \frac{m_q^2}{s} \: \alpha_g \right ) \: \Theta (\beta_q -
\beta_g); \nonumber \\
& & \nonumber \\
C_{10{\rm b}} & = & C_F \left ( C_F \: - \: \frac{C_A}{2} \right ) \:
\Theta (\alpha_q - \alpha_g) \: \Theta (\beta_g - \beta_q) \:
\Theta \left (\alpha_g \: - \: \frac{m_q^2}{s} \: \beta_g \right
); \nonumber \\
& & \nonumber \\
C_{10{\rm c}} & = & C_F \frac{C_A}{2} \: \Theta (\beta_g - \beta_q) \:
\Theta \left (\alpha_g \: - \: \frac{m_q^2}{s} \: \beta_g \right
); \nonumber \\
& & \nonumber \\
C_{10{\rm d}} & = & C_F \left (C_F \: - \: \frac{C_A}{2} \right ) \:
\Theta (\beta_g - \beta_q) \: \Theta \left (\alpha_g \: - \:
\frac{m_q^2}{s} \: \beta_g \right ); \nonumber \\
& & \nonumber \\
C_{10{\rm e}} & = & - C_F \left (C_F \: - \: \frac{C_A}{2} \right ) \:
\Theta (\beta_g - \beta_q) \: \Theta (\alpha_q - \alpha_g) \:
\Theta \left (\alpha_g \: - \: \frac{m_q^2}{s} \: \beta_g \right
); \nonumber \\
& & \nonumber \\
C_{10{\rm f}} & = & C_F \frac{C_A}{2} \: \Theta (\alpha_g - \alpha_q)
\: \Theta (\beta_g - \beta_q); \\
& & \nonumber \\
C_{10{\rm g}} & = & 0. \nonumber
\eea
Using these results it is straightforward to show that the total DL contribution
of Fig.~10 is given by (\ref{eq:f5}).

\newpage
\noindent {\large \bf Figure Captions}
\begin{itemize}
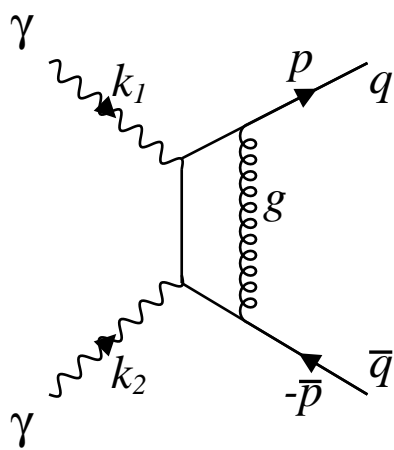
\item[Fig.~1] A virtual contribution to the background process
$\gamma\gamma 
\rightarrow q \overline{q}$.  There is also a contribution with
$k_1 
\leftrightarrow k_2$. These are the only one-loop diagrams which give
non-vanishing contributions in the $m_q = 0$ limit if the helicities of the 
photons are equal.

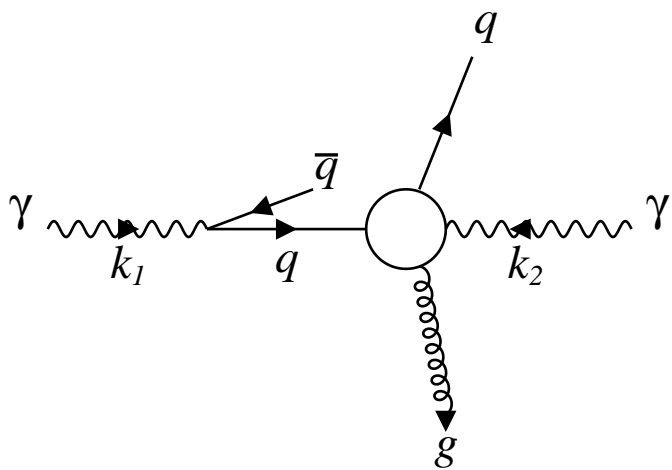
\item[Fig.~2] The Compton configuration of $\gamma\gamma
\rightarrow 
q\overline{q}g$ which can fake the Higgs signal if only one
energetic $b$ 
jet is tagged.  There are also contributions with $q
\leftrightarrow \overline{q}$ 
and/or $k_1 \leftrightarrow k_2$.

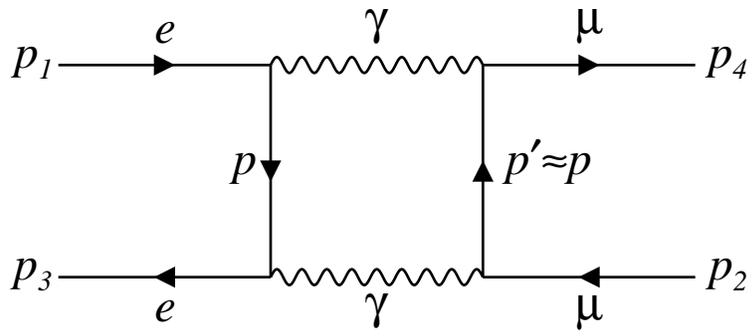
\item[Fig.~3] The diagram which gives a DL contribution to
backward 
electron-muon scattering.

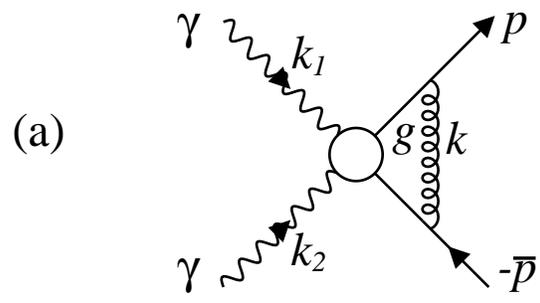
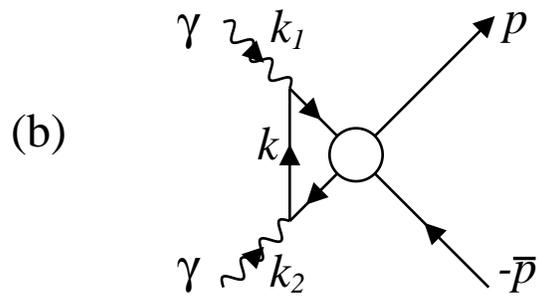
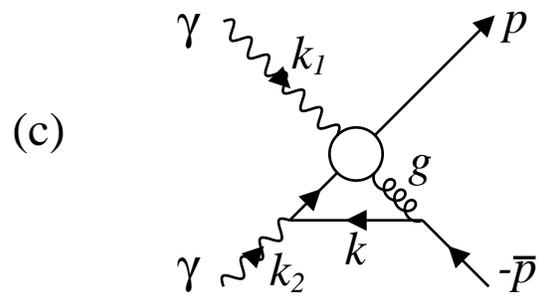
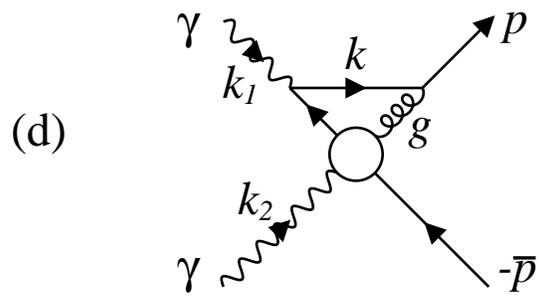
\item[Fig.~4] Four (non-overlapping) configurations which give
${\cal O} (\alpha_S)$ DL corrections to $\gamma\gamma \rightarrow
q\overline{q}$.  For diagrams (b), (c) and (d) there are also
contributions with $k_1 \leftrightarrow k_2$.

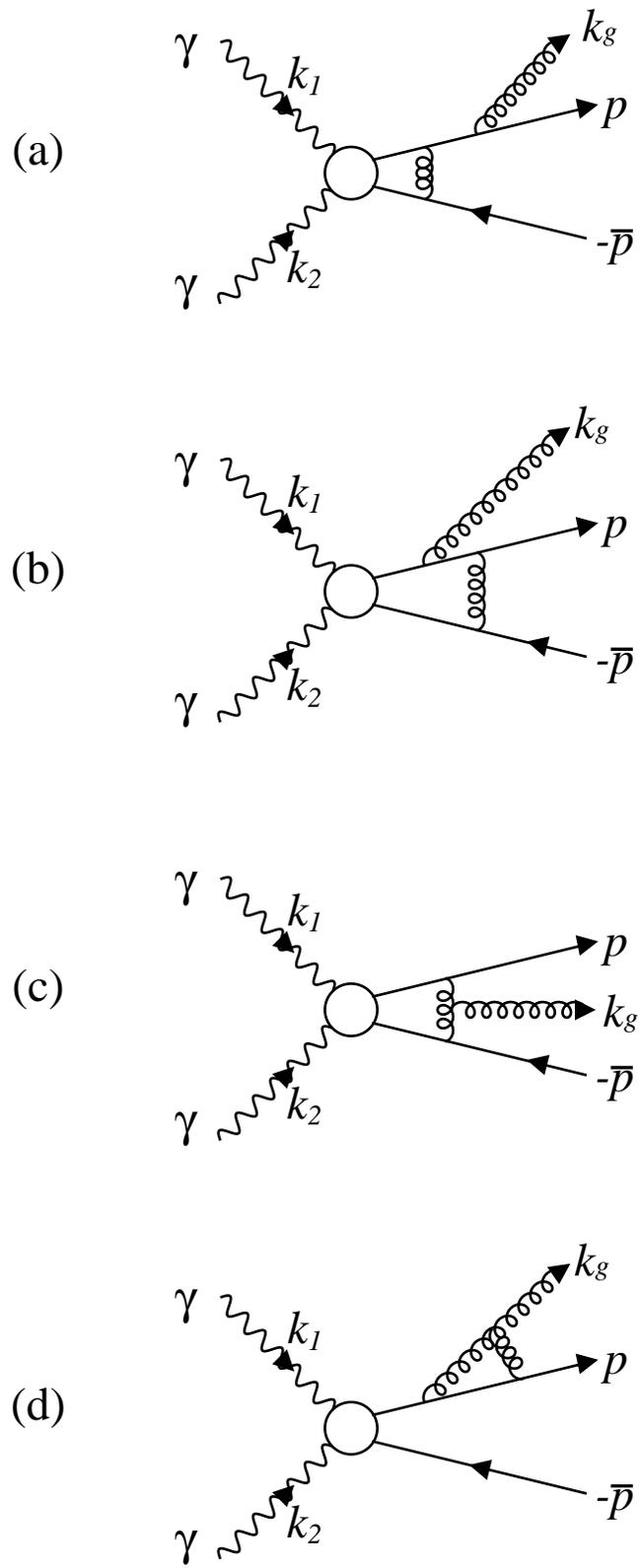
\item[Fig.~5] The soft gluon emissions in the case when the hard subprocess is
$\gamma \gamma \rightarrow \ q \overline{q}$.  We also have diagrams (labelled
$5 ({\rm \overline{a}, \overline{b}, \overline{d}})$ in the text) in which the gluon
is emitted from the $\overline{q}$ rather than the $q$.  In addition all 
these diagrams have counterparts with $k_1 \leftrightarrow k_2$.

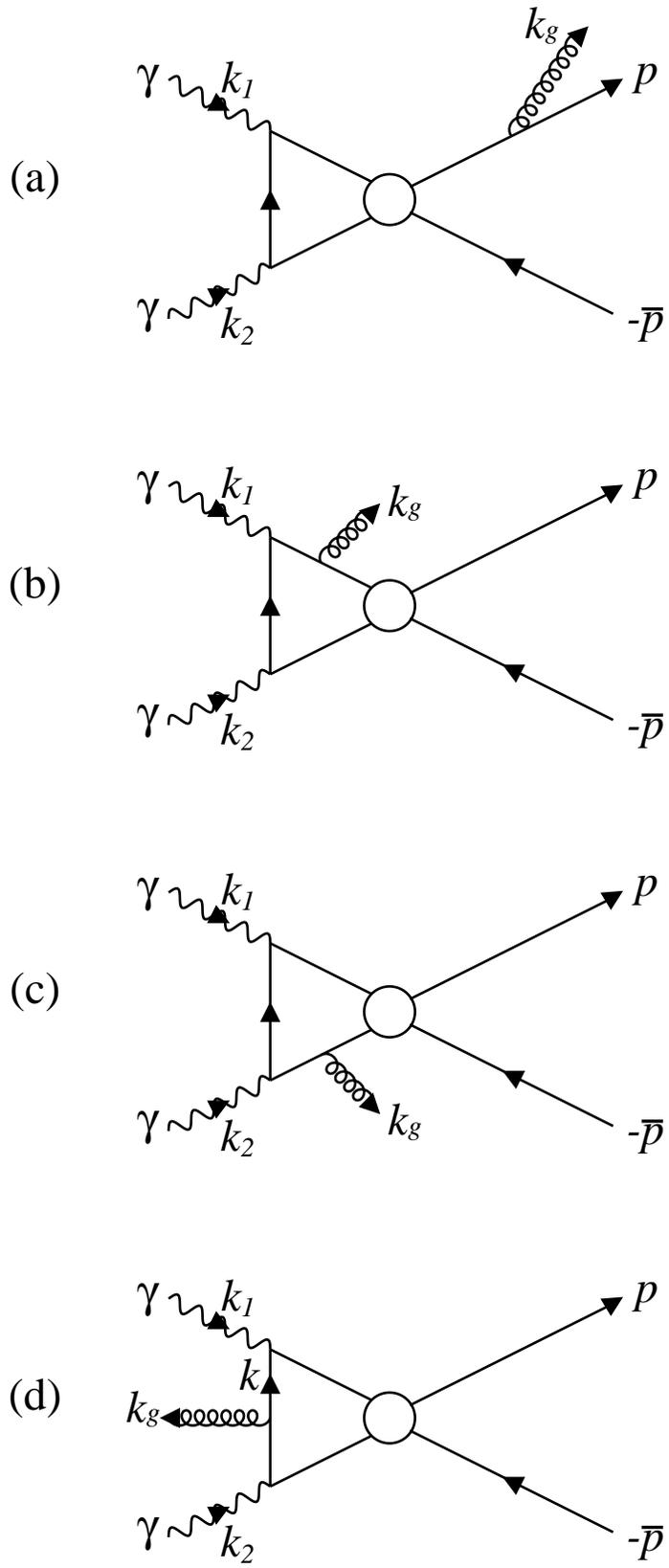
\item[Fig.~6] The soft gluon emission from the diagram shown in
Fig.~4(b).  There is also a diagram (labelled $6({\rm \overline{a}}$) in the text) where
the gluon is emitted from the $\overline{q}$ rather than the $q$.  In addition
all these diagrams have counterparts with $k_1 \leftrightarrow k_2$.

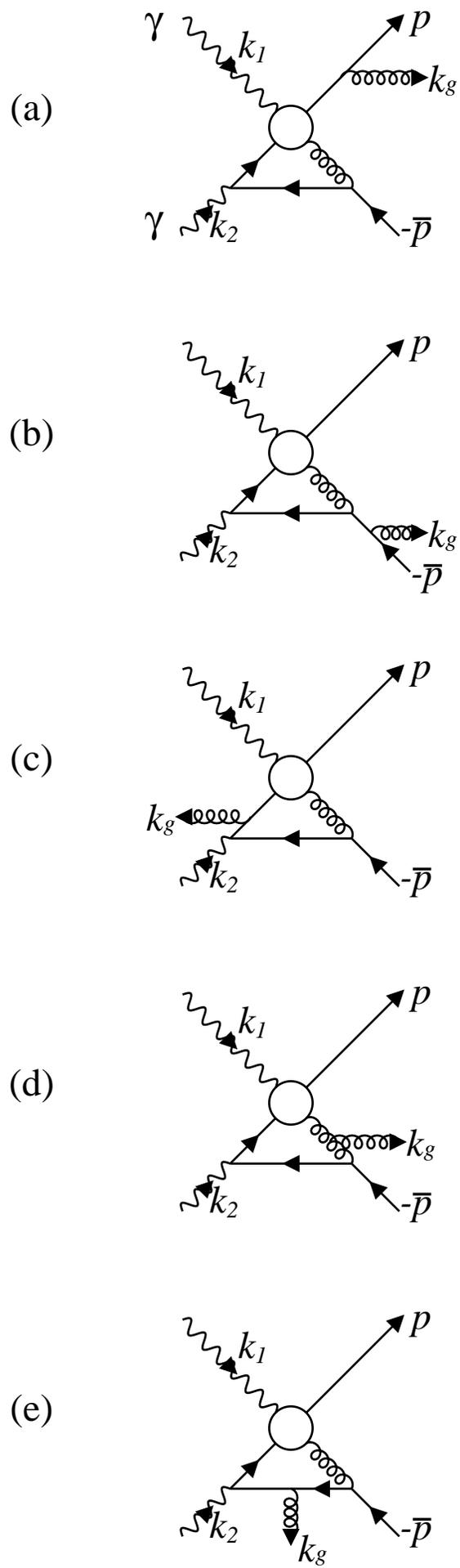
\item[Fig.~7] The soft gluon emission from the diagram shown in
Fig.~4(c).  There are also contributions with $k_1 \leftrightarrow
k_2$.

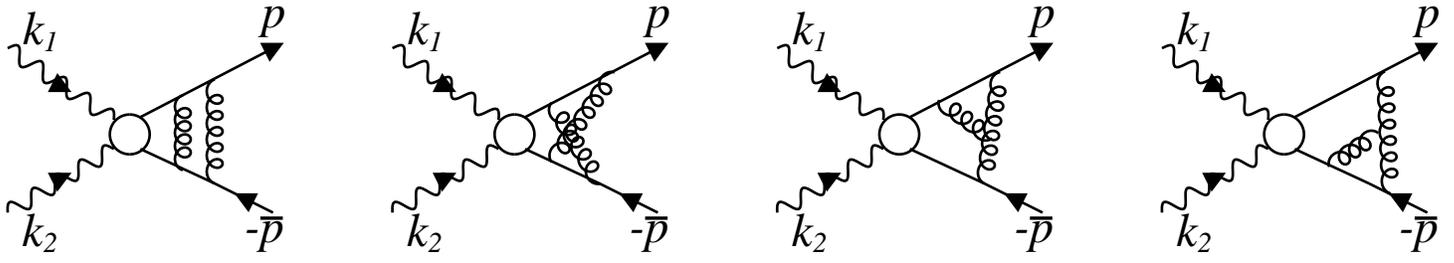
\item[Fig.~8] Diagrams for the non-radiative 2-loop corrections when the
hard subprocess is $\gamma \gamma \rightarrow q \overline{q}$.

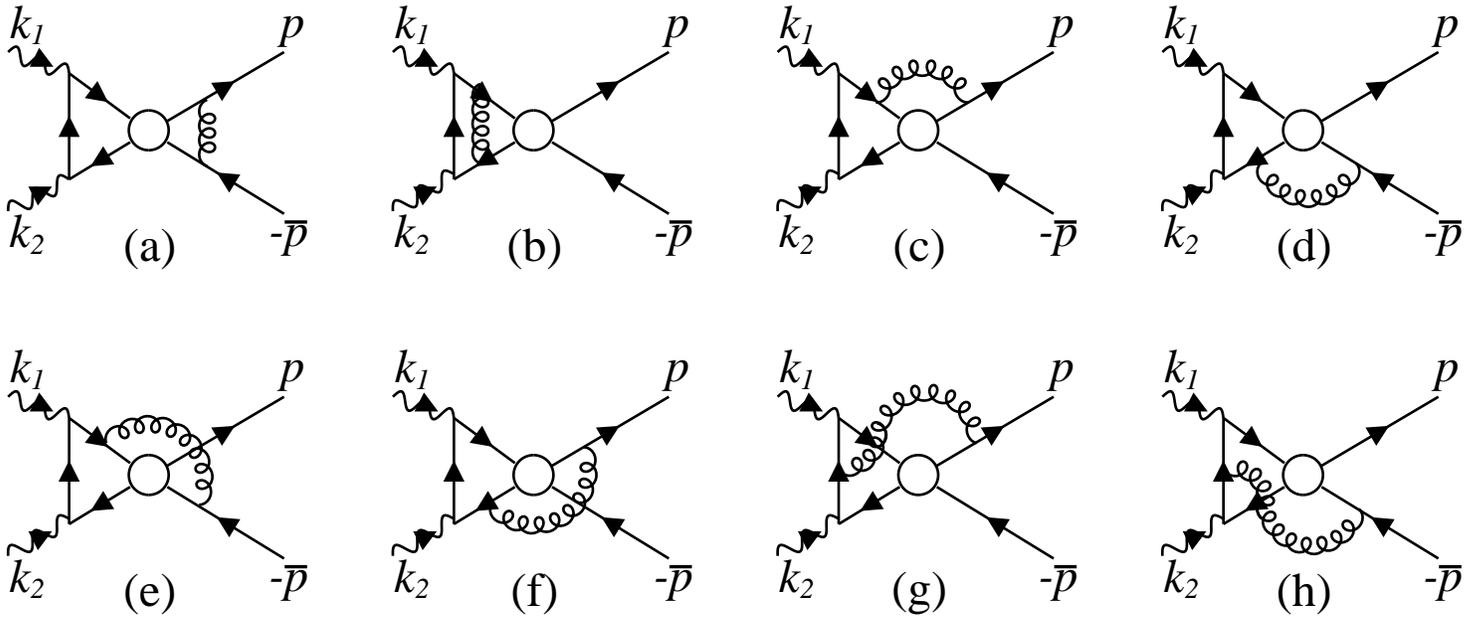
\item[Fig.~9] Diagrams for the non-radiative 2-loop corrections when the hard
subprocess is $q \overline{q} \rightarrow q \overline{q}$.  There are also
contributions with $k_1 \leftrightarrow k_2$.

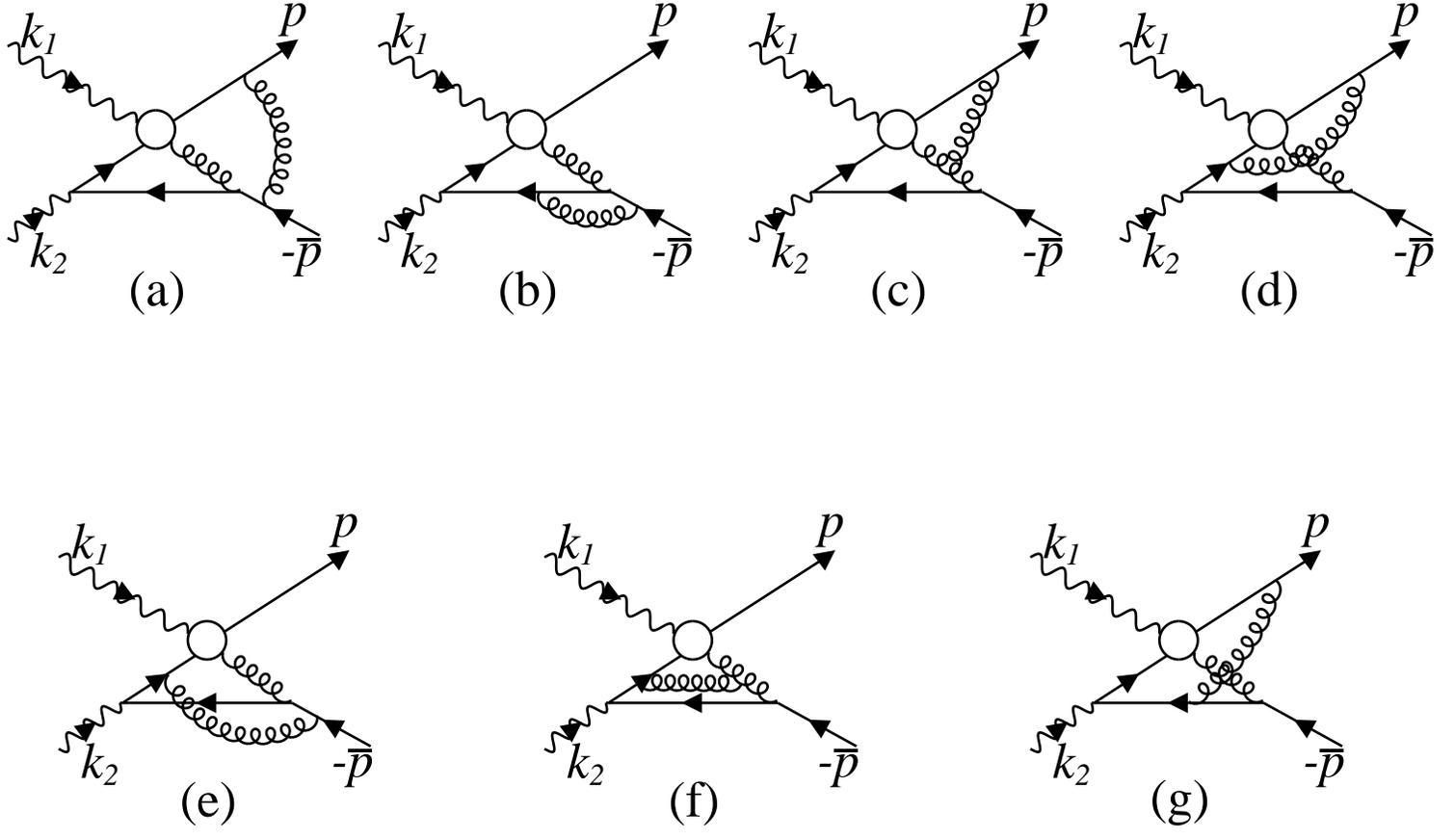
\item[Fig.~10] Diagrams for the non-radiative 2-loop corrections when the hard
subprocess is $\gamma q \rightarrow q g$.  There are also contributions with
$k_1 \leftrightarrow k_2$.

\end{itemize}

\end{document}